\documentclass[12pt]{article}

\usepackage{float}
\usepackage{amsthm}
\usepackage{amsmath}
\usepackage{amssymb}
\usepackage[top=2.7cm, bottom=3cm, left=2.7cm, right=2.7cm]{geometry}
\usepackage{graphicx,subfigure}
\usepackage{mathrsfs}
\usepackage{multirow}
\usepackage{natbib}

\usepackage[oldcommands]{algorithm2e} % after natbib!
\usepackage{paralist}
\usepackage[toc,page,title,titletoc,header]{appendix}
\usepackage{booktabs}  % \cmidrule[width](trim){a–b}  \cmidrule(r){1-2}
\usepackage[colorlinks,linkcolor=blue,anchorcolor=black,citecolor=blue]{hyperref}

\usepackage{authblk}

\usepackage{setspace}
\newtheorem{definition}{Definition}
\newtheorem{theorem}{Theorem}
\newtheorem{lemma}{Lemma}
\theoremstyle{remark}

\newcommand{\bsb}{\boldsymbol}
\newcommand{\tr}{\mbox{tr}}

\newcommand{\bsbA}{{\boldsymbol{A}}}
\newcommand{\bsbB}{{\boldsymbol{B}}}
\newcommand{\bsbC}{{\boldsymbol{C}}}
\newcommand{\bsbD}{{\boldsymbol{D}}}
\newcommand{\bsbE}{{\boldsymbol{E}}}

\newcommand{\bsbG}{{\boldsymbol{G}}}

\newcommand{\bsbI}{{\boldsymbol{I}}}
\newcommand{\bsbJ}{{\boldsymbol{J}}}

\newcommand{\bsbL}{{\boldsymbol{L}}}

\newcommand{\bsbP}{{\boldsymbol{P}}}

\newcommand{\bsbR}{{\boldsymbol{R}}}
\newcommand{\bsbS}{{\boldsymbol{S}}}

\newcommand{\bsbU}{{\boldsymbol{U}}}
\newcommand{\bsbV}{{\boldsymbol{V}}}
\newcommand{\bsbW}{{\boldsymbol{W}}}
\newcommand{\bsbX}{{\boldsymbol{X}}}
\newcommand{\bsbY}{{\boldsymbol{Y}}}
\newcommand{\bsbZ}{{\boldsymbol{Z}}}

\newcommand{\bsbs}{{\boldsymbol{s}}}
\newcommand{\bsbu}{{\boldsymbol{u}}}
\newcommand{\bsbv}{{\boldsymbol{v}}}

\newcommand{\bsbx}{{\boldsymbol{x}}}
\newcommand{\bsby}{{\boldsymbol{y}}}

\newcommand{\bsba}{{\boldsymbol{\alpha}}}
\newcommand{\bsbbeta}{{\boldsymbol{\beta}}}

\newcommand{\bsbmu}{{\boldsymbol{\mu}}}

\newcommand{\bsbVp}{{\bsbV_{\perp}}}
\newcommand{\bsbVpt}{{\bsbV_{\perp}^T}}
\newcommand{\bsbtX}{{\boldsymbol{\tilde{X}}}}

\newcommand{\bsbhatVp}{{\hat{\bsbV}_{\perp}}}
\newcommand{\bsbtV}{{\boldsymbol{\widetilde{V}}}}

\newcommand{\bsbDelta}{{\boldsymbol{\Delta}}}
\newcommand{\bsbxi}{{\boldsymbol{\xi}}}

\newcommand{\Proj}{{\mathbf P}}
\newcommand{\rd}{\,\mathrm{d}}
\newcommand{\EP}{\,\mathbb{P}}
\newcommand{\EE}{\,\mathbb{E}}

\DeclareMathOperator{\vect}{\mbox{vec}\,}
\providecommand{\keywords}[1]{\textbf{Keywords: } #1}

\begin{document}

%%%%% Short version %%%%%%
%%%% for jasa only %%%
%\setlength{\abovedisplayskip}{0mm plus2mm minus2mm}
%\setlength{\belowdisplayskip}{0mm plus2mm minus2mm}
%%\setlength{\arraycolsep}{0cm plus4mm minus3mm}
%
%\setlength{\topsep}{0mm plus2mm minus2mm}
%\setlength{\partopsep}{0mm plus2mm minus2mm}
%\setlength{\itemsep}{0mm plus2mm minus2mm}
%
%\setlength{\abovecaptionskip}{0mm plus2mm minus2mm}
%\setlength{\belowcaptionskip}{0mm plus2mm minus2mm}
%%%%%%%%%%%%
%%%%% Short version ends %%%%%%

\title{Robust Orthogonal Complement Principal Component Analysis}
%\author{\small Yiyuan She, Shijie Li and Dapeng Wu}
%\author{\small}
\author[1]{Yiyuan She}
\author[2]{Shijie Li and Dapeng Wu}
\affil[1]{Department of Statistics, Florida State University}
\affil[2]{Department of Electrical \& Computer Engineering, University of Florida}

\date{}
\maketitle

%\onehalfspacing
%
%{\small{
%Yiyuan She is Associate Professor, Department of Statistics, Florida State University, Tallahassee, FL 32306 (yshe@stat.fsu.edu).
%Shijie Li received Ph.D. degree in Electrical and Computer Engineering from University of Florida in 2014 (lishijie0602@gmail.com).
%Dapeng Wu is Professor, Department of Electrical \& Computer Engineering, University of Florida,  Gainesville, FL 32611 (wu@ece.ufl.edu).  This work was supported in part by NSF grants  CCF-1117012, CCF-1116447 and DMS-1352259.
% We would like to thank the editor, the associate editor and two anonymous referees
%for their careful comments and useful suggestions that significantly
%improve the quality of the paper.
%}}

\begin{abstract}
Recently, the robustification of principal component analysis has attracted lots of attention from statisticians, engineers and computer scientists. In this work we study the type of outliers that are not necessarily apparent in the original observation space but can seriously affect the principal subspace estimation. Based on a mathematical formulation of such transformed outliers, a novel robust orthogonal complement principal component analysis (ROC-PCA) is proposed. The framework combines the popular sparsity-enforcing and low rank regularization techniques to deal with row-wise outliers as well as element-wise outliers.
A non-asymptotic oracle inequality  guarantees the accuracy and high breakdown   performance of ROC-PCA in finite samples.
To tackle the computational challenges,  an efficient algorithm is developed on the basis of Stiefel manifold optimization and iterative thresholding. Furthermore, a batch variant is proposed to significantly reduce the  cost in ultra high dimensions. The paper also  points out a   pitfall of a common practice of  SVD  reduction   in robust PCA.  Experiments show the effectiveness and efficiency of ROC-PCA in both synthetic and real data.
%This article has supplementary material online.
\end{abstract}


%\setstretch{1.6}
%\doublespacing

\section{Introduction}
\label{sec:introduction_rocpca} During the past few years, big data arising in machine learning, signal processing, genetics, and many other fields pose a dimensionality challenge in statistical computation and analysis. To uncover low-dimensional structures underlying such high-dimensional data, the {principal component analysis} (PCA) %~\citep{hotelling1933analysis}
is one of the most popularly used multivariate dimension reduction tools.
Let $\bsbX \in \mathbb{R}^{n\times p}$ be a data matrix with $n$ observations in $p$-dimensional space. PCA can be characterized by finding a low rank data approximation, i.e.,  $\min_{\bsbB}$ $\|\bsb{X-B}\|_F^2$ subject to $\text{rank}(\bsbB)\le r$.
% \begin{equation}
% \label{eq:matrixapproximation}
% \min_{\bsbB} \|\bsb{X-B}\|_F^2 \quad \text{s.t.} \quad \text{rank}(\bsbB)\le r.
% \end{equation}
The solution is given by a truncated singular value decomposition (SVD) of $\bsbX$: $\hat{\bsbB}=\bsbU\mbox{diag}\{\sigma_1,...,\sigma_r\}\bsbV^T=\bsb{XVV}^T$, where $\bsbV$ consists of the first $r$ right singular vectors of $\bsbX$ with $\bsb{VV}^T$ defining the rank-$r$ principal subspace. The squared error loss function is reasonable under  a Gaussian noise model $\bsbX=\bsbB+\bsbE$, but is notoriously known to be non-robust and sensitive to atypical observations or the so-called \emph{outliers}. Outliers typically refer to extreme observations far away from the majority of the data, and occur ubiquitously in real life data  (\cite{maronna2006robust}, \cite{hampel2011robust}). They may seriously affect statistical estimation and inference---in fact, a single outlier can break down the PCA completely and result in a misleading subspace estimate.

The robustification of PCA has been extensively studied in robust statistics, e.g., \cite{rousseeuw1999fast}, \cite{locantore1999robust}, \cite{hubert2005robpca}, among many others. The recent renowned   \emph{Principal Component Pursuit} (PCP) due to \cite{Candes2011} has drawn a lot of attention from researchers even beyond the statistics community. PCP decomposes $\bsbX$ into a low-rank component $\bsbB$ and a sparse gross outlier component $\bsb{S}$. The recovery problem can be formulated by $\min_{\bsbB,\bsb{S}} \text{rank}(\bsbB)+\lambda\|\bsb{S}\|_0$ subject to $\bsb{X}=\bsbB+\bsb{S}$, where $\|\cdot\|_0$ denotes the element-wise $\ell_0$ norm, i.e., the number of all non-zeros. PCP applies a convex relaxation  to facilitate computation and analysis:
$\min_{\text{$\bsbB$, $\bsb{S}$}} \|\bsbB\|_* + \lambda\|\bsb{S}\|_1 \ \   \text{subject to} \ \ \bsb{X}=\bsbB+\bsb{S},
$ % absolute values of all entries
where $\|\cdot\|_*$ denotes the matrix {nuclear norm} (sum of all singular values), and $\|\cdot\|_1$ denotes
the element-wise $\ell_1$ norm. PCP has various extensions and variants (\cite{zhou2010stable}, %\cite{ganesh2010dense},
\cite{xu2010robust}, \cite{wright2013compressive}), and has widespread applications in image and video analysis, e.g., \cite{wright2009robust}, \cite{peng2012rasl}, \cite{zhang2012tilt}. Although PCP can effectively deal with additive outliers in the original observation space, it may fail in the presence of another important type of outliers, the so-called \emph{OC outliers}, which is the major concern of this work.

\begin{figure}[!htbp]
    \centering
    \subfigure[Normal data ('.') and PC outliers ('+').]{
    \label{fig:PCoutlier}
    \includegraphics[height=0.31\textwidth,width=0.47\textwidth,clip]{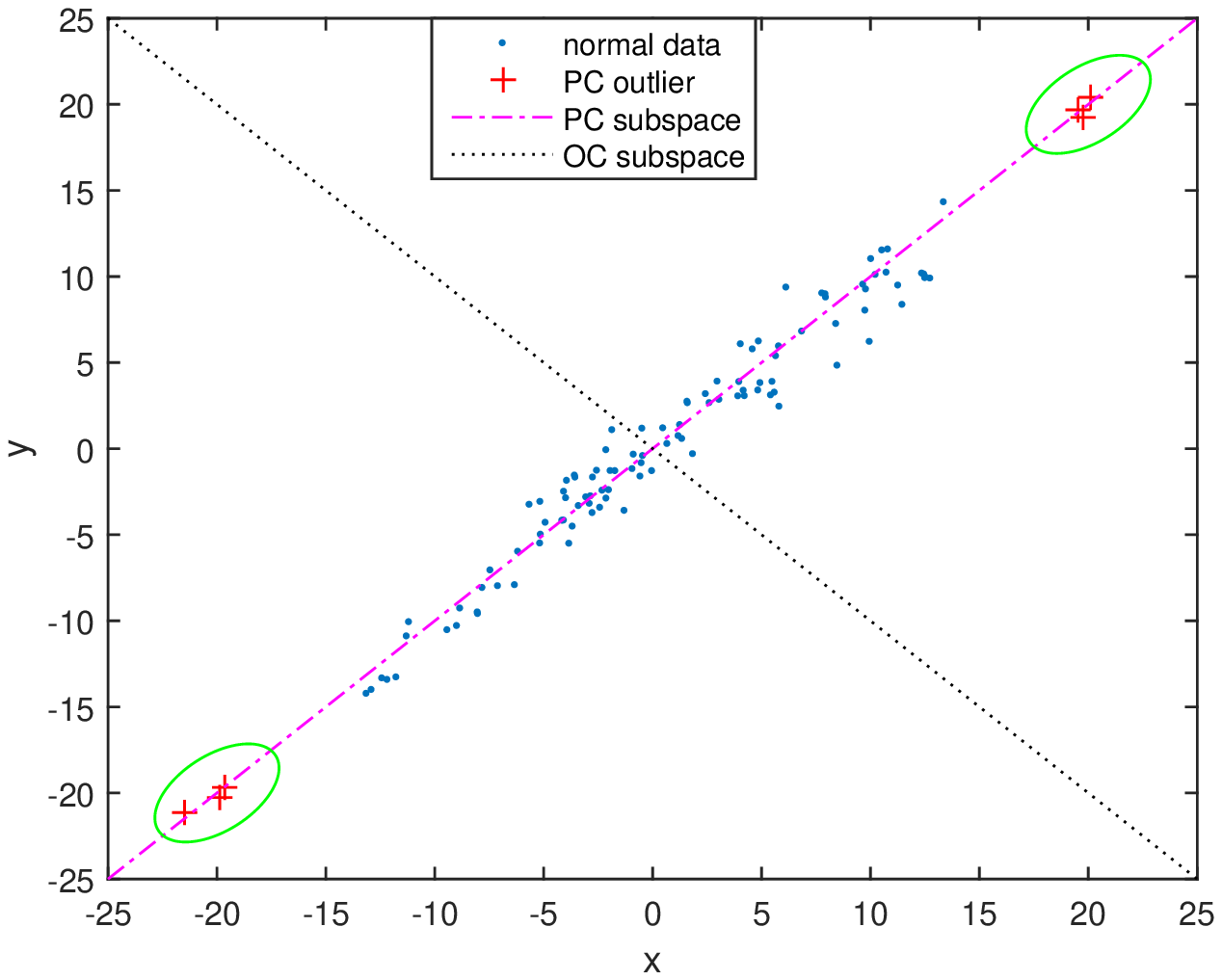}
    }
    \subfigure[Normal data ('.') and OC outliers ('o').]{
    \label{fig:OCoutlier}
    \includegraphics[height=0.31\textwidth,width=0.47\textwidth,clip]{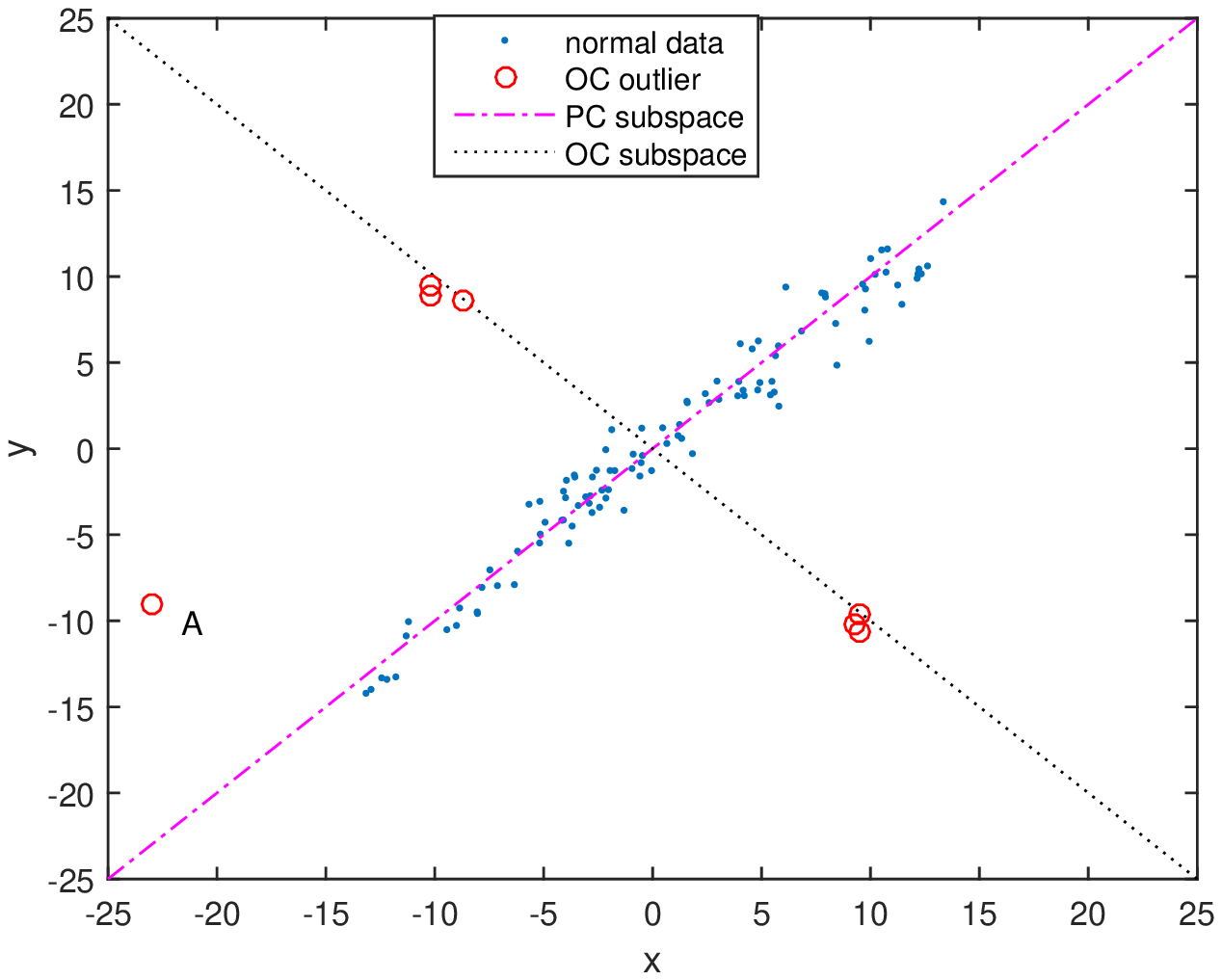}
}
    \caption{PC outliers and OC outliers.}
    \label{fig:motivation}
\end{figure}

In robust principal component analysis, the outliers worthy of attention must affect the principal subspace estimation. Figure~\ref{fig:motivation} gives some toy examples to illustrate how outliers could interfere with principal subspace identification. In the left panel, some outlying samples exist in the principal component subspace (\textbf{PC subspace}), which we call \textit{pure} \textbf{PC outliers}, or PC outliers, for short. Interestingly, they do not affect the detection of PC subspace.  %That is, for the purpose of principal subspace estimation,
 PC outliers are not that harmful (though possibly affecting the order of   PC  directions), and thus, can be handled in a later stage.  However, if one only checks the raw $(x$, $y)$ coordinates in the observed space, these samples might be labeled as outliers.

The right panel contains some  samples atypical in the orthogonal complement subspace (\textbf{OC subspace}), which we call \textbf{OC outliers}. We emphasize that \textit{any} points showing  outlyingness in  OC coordinates are referred to as  OC outliers in this paper, such as Point $A$ in Figure  \ref{fig:OCoutlier}, whether or not they show PC outlyingness. It is the OC outliers that can skew the PC subspace.
 Unfortunately, checking their coordinates in the observation space offers little help, thereby making PCP  possibly fail in recovering the genuine PC subspace. However, if one could project the data points onto the ideal OC subspace, such outliers would be easily revealed and detected.

This paper proposes a novel robust orthogonal complement principal component analysis (ROC-PCA) to address such OC outliers in principal subspace recovery. In contrast to the existing robust PCA approaches, ROC-PCA explicitly deals with  OC outliers, and aims at simultaneous outlier identification and robust principal subspace recovery. Both \emph{row-wise} (r-Type) outliers and \emph{element-wise} (e-Type) outliers are discussed. Our computation algorithm involves Stiefel manifold optimization and allows for all popular sparsity-enforcing penalties to be used.
%In addition to showing the high breakdown point of ROC-PCA,
We also establish a non-asymptotic oracle inequality to provide a theoretical guarantee of ROC-PCA from the predictive learning perspective.

The rest of the paper is organized as follows.
%%%% Long version %%%%%
Section~\ref{sec:survey_rocpca} briefly reviews the existing robust PCA methods and models.
%%%% Long version ends %%%%
Section~\ref{sec:formulation_rocpca} proposes the ROC-PCA and formulates a useful framework to generalize the $M$-estimators to robust subspace estimation. In Section~\ref{sec:computation_rocpca}, a class of computational algorithms involving Stiefel manifold optimization and iterative thresholding is proposed. Section~\ref{sec:theory_rocpca} theoretically analyzes the performance of ROC-PCA in finite samples. In Section~\ref{sec:srocpca_rocpca}, we point out a pitfall of applying a popular SVD reduction in high dimensional robust PCA and  propose a \emph{batch} variant of ROC-PCA for big data computation. Section \ref{sec:simulation_rocpca} and Section \ref{sec:realdata}
present %real data analysis.
extensive numerical studies using both synthetic data and real data, and compare ROC-PCA with some other popular robust PCA approaches.
We conclude  in Section~\ref{sec:conclusion_rocpca}.
%%%%% Long version %%%%%
All technical details are left to the Appendices.
%%%%% Long version ends %%%
%%%%%%% Short version %%%%
%A survey of   robust PCA methods and models, an algorithm summary,  simulation studies, and all technical details are left to the supplementary material.
%%%%%% Short version ends %%%
\section{Survey of Robust PCA}
\label{sec:survey_rocpca}

The non-robustness issue of PCA has been noticed long before and  extensively investigated in robust statistics. See, e.g.,  \cite{maronna2006robust}\ for a comprehensive introduction. In this work, we classify  robust PCA approaches into five classes: 1) \emph{robust covariance matrix} based methods; 2) \emph{projection} based methods; 3) \emph{hybrid} projection-covariance estimation based methods; 4) \emph{spherical/elliptical} PCA; 5) \emph{low rank matrix approximation} based methods.  Due to space limitations, we only review some representative works in each class, which gives by no means a complete list of literature on robust PCA.
 % In this section, we mainly review the methods that can handle OC outliers.

\paragraph{Robust covariance matrix based methods}
Robust PCA can be achieved by first finding a robust covariance estimate, and then extracting PC loadings from the eigen-decomposition of this matrix. One of the earliest works  is \cite{maronna1976robust} that considers   multivariate $M$-estimation of location and dispersion with monotone $\psi$-functions.
%Huber's $\psi$-function can be used, however, its breakdown point is at most $1/(p+1)$.
 %\cite{croux2000principal} recommend
 Using the $S$-estimator~\citep{davies1987asymptotic} as a scale measurement can  gain high breakdown, but it leads to large estimation bias  when $p$ is large~\citep{rocke1996robustness}.
% The earliest work in this class is~\cite{maronna1976robust}, where a weighted covariance matrix is obtained by means of $M$-estimators.Yet this method cannot tolerate multiple outliers, and
% In fact, it is an iterative updating procedure to solve a set of implicit equations. In the procedure, a weighted covariance matrix with data-dependent weights is obtained via a weight function of the Mahalanobis distance for each observation, which essentially measures the outlyingness of that observation. Specifically, full weights are given to those samples from the bulk of the data, while reduced weights are given to the outlying samples.
\cite{rousseeuw1985multivariate} proposed the Minimum Volume Ellipsoid (MVE) and  Minimum Covariance Determinant (MCD) for robust covariance matrix estimation, where   data resampling  is effectively used to speed the computation. Given $n$ data points, by repeatedly resampling subsets of size $h$ ($n/2\le h <n$), an optimal subsample can be obtained according to the MVE or MCD criterion;  the sample covariance matrix from this subset is then  delivered.
% such that certain criteria are optimized
% For the MVE, it tries to find the $h$ subset that expands the ellipsoid with smallest volume, which, in essence, is equivalent to minimizing the median of the Mahalanobis distances between the observations and the robust mean vector under the robust covariance deflation \citep{maronna2006robust}. In contrast, the MCD looks for the $h$ subset such that the covariance matrix of the data inside this subset has the lowest determinant, which reduces to minimizing a trimmed scale estimate of the Mahalanobis distances between the
% observations and the robust mean vector under the robust covariance deflation \citep{maronna2006robust}.
Both methods have high breakdown points, but MCD is more efficient,  in terms of both computation and statistical estimation~\citep{davies1992asymptotics}. The fast-MCD algorithm  \citep{rousseeuw1999fast} runs more efficiently, largely owing to a concentration step to keep the cost function value decreasing throughout the iterations.

Some limitations of this class of approaches are as follows. 1) The computational cost is relatively high, especially for large-$p$ data. In our experience, even the fast-MCD becomes computationally intractable when $p>100$. 2) The aforementioned methods do not directly apply to data matrices with rank deficiency, such as modern high dimensional datasets with $p>n$. For example, the determinant of the sample covariance matrix collapses to zero in MCD. 3) They cannot accommodate element-wise outliers. In this scenario, the portion of atypical observation rows can easily go beyond $50\%$, which contradicts the assumption in subset sampling. 4) Robust subspace estimation and inherent outlier detection cannot be achieved simultaneously. To identify the outliers, a cut-off value has to be chosen, which is not a trivial task. It is also worth mentioning that for the purpose of rank-$r$ PC subspace estimation, estimating the whole covariance matrix (robustly) might be unnecessary.

\paragraph{Projection based methods}
The second class is projection based, to reduce the multivariate problem to many univariate ones. %The projection idea is related to ~\cite{stahel1981breakdown} and~\cite{donoho1982breakdown}.
In general, this class of methods repeatedly conducts rank-1 robust PCA, seeking the direction which maximizes a robust  scale measure of the projected data. Usually the data matrix is deflated  once a new direction is obtained. % (see, e.g., \cite{mackey2008deflation}). %$M$-estimator in evaluating the scale

\cite{li1985projection}   formulated the idea systematically  and showed  some robustness properties of the  estimation procedures. Their
%Huber's concomitant scale estimator.
 projection pursuit based algorithm is however computer-intensive. The C-R algorithm \citep{croux1996fast} is more computationally attractive in that the search of a new PC direction is restricted to a set of trial directions determined  by the data center and every data point.  The $Q_n$ scale estimator \citep{rousseeuw1993alternatives} is recommended to achieve good efficiency and high breakdown \citep{croux2005high}. Noticing that the C-R algorithm  suffers from accumulated round-off errors, \cite{hubert2002fast} proposed the RAPCA  made up of  an SVD reduction step and a reflection step. The SVD reduction is detailed in  Section \ref{sec:srocpca_rocpca} in the text---unfortunately, it may have serious issues  in high dimensions.  The reflection step is a Householder transformation based deflation,   conducted each time a new PC direction is found.

Projection based methods bypass the estimation of the whole covariance matrix and can be applied to data matrices with rank deficiency. Most of the methods, such as the C-R algorithm, are sequential and run faster than those in the first class \citep{croux2005high}. On the other hand, they may not meet all application needs. 1) Without the help of SVD reduction (which may  be un-trustworthy for large $p$), they can only be used in moderately high dimensions.
%When $p$ is ultra-high, there are too many candidate directions to choose from.
Enumerating all `trial directions', though affordable,  may not cover all candidate PC directions in the $p$-dimensional space.  (Indeed, the number of such  directions  only relies on $n$ no matter how large $p$ is.)  2) The sequentially obtained PC directions may not guarantee joint optimality. (Note that the loss of joint orthogonality is, however, acceptable in many applications). The design of deflation can also be  quite \emph{ad hoc}, see, e.g.,  \cite{mackey2008deflation}. 3) Similar to the first class, they can not deal with element-wise outliers.
% \textbf{There is no evidence showing which class of methods is faster, and each class contains many methods, it is hard to say which class is better, so I just ignored the comparisons.}
%The computational cost is lower in comparison with the robust covariance matrix based methods.

\paragraph{Hybrid projection-covariance estimation based methods} % ($h<n$)
A representative work in this class is the ROBPCA~\citep{hubert2005robpca}. It is an involved multiple-step procedure.  First, perform an SVD reduction to project raw data points onto the observed space of lower dimensionality. Next, find a subset containing $h$ {least outlying} observations. The outlyingness measure \citep{stahel1981breakdown,donoho1982breakdown} is projection based; \textbf{only}  the  directions passing through  two data points are sampled (250 times).  Then, apply  fast-MCD  to this $h$-subset.  %(with another SVD reduction conducted to eliminate rank deficiency).
 Finally, to further increase the estimation efficiency, a reweighted covariance matrix is computed based on the mean and covariance estimates from the fast-MCD.

ROBPCA shares the high breakdown property of MCD and shows excellent performance in many of our simulation experiments. Yet it also suffers from some aforementioned drawbacks of  projection or robust covariance based methods. For instance, ROBPCA cannot handle element-wise anomalies,   the fast-MCD step  may not be efficient enough  in big data computation, and the SVD reduction may be problematic when $n\ll p$ (cf. Section \ref{sec:srocpca_rocpca} in the text).
Another problem lies in the sampling restriction. It is not difficult to see that  to evaluate the Stahel-Donoho outlyingness accurately,  the trial directions are better  in  the OC subspace. (This is in contrast to the C-R algorithm in which the purpose of   trial directions is to cover the PC directions.) ROBPCA may fail in some very simple setups because of this restriction (Section~\ref{sec:simmulationComparison_rocpca}).
\paragraph{Spherical/elliptical PCA}
The scheme for spherical/elliptical PCA~\citep{locantore1999robust} %,marden1999some}
is relatively simple: project all data points (centered) onto a unit sphere or an ellipse, and  apply the (plain) PCA to the spherical/elliptical data.

Spherical PCA is convenient and fast even in high dimensions. The choice of the data center is crucial. It is perhaps preferable to estimate the center and recover the PC subspace simultaneously in this multivariate setup (in the spirit of \eqref{eq:MStransformed}, for instance). The projection step   preserves  angle information but loses all distance information. In certain applications this may result in  ambiguities or  misleading subspace estimates. Consider a toy example in $\mathbb R^2$. Two data points are at $(\pm 1$, $0)$, and the rest $(n-2)$ points are at $(\varepsilon\cdot\cot{\frac{2\pi}{n}k}$, $\pm \varepsilon)$, $k=1, \cdots, \frac{n}{2}-1$, i.e., the intersections of two horizontal lines $y=\pm \varepsilon$ and the radii with polar angles $\pm \frac{2\pi}{n}k$, $k=1,\cdots,\frac{n}{2}-1$. When $\varepsilon$ is small and $n$ is large, the $x$-axis should give the dominant PC direction. However, spherical PCA favors none of these radius directions. Even worse, with two more data points  at $(0$, $\pm \varepsilon)$, spherical PCA would mistakenly take the $y$-axis as the PC direction. From the experiment results in Section \ref{sec:simulation_rocpca}, this method has limited power when the number of outliers is relatively large.  %Statistically, \cite{serneels2006spatial} show that the projection step, which amounts to a spatial sign transformation of the data, can lead to a very biased subspace estimate.

\paragraph{Low rank matrix approximation based methods}
This class of methods uses low rank data approximation  and minimizes a  robust  fitting criterion. For example,  \cite{croux2003fitting} designed a weighted $\ell_1$ loss function with data-dependent weights and  \cite{maronna2008robust} recommended using resistant error loss functions associated with redescending $\psi$ functions. In addition, it is easy to show that the stable (noisy)  PCP
%~\citep{Candes2011}
\citep{zhou2010stable}  amounts to   applying the Huber's loss---see \cite{she2011outlier} for a general connection between penalty functions and robust loss functions.
Another interesting convex relation is given by \cite{zhang2014novel}. %A combination of ideas of  projection and robust data approximation can be found in  \cite{l1-proj}.

This class of methods usually copes with element-wise outliers in the original observation space \citep{Candes2011}, but may not be able to deal with OC outliers effectively. On the other hand, PCP serves as one important motivation for our ROC-PCA.

% one can use the $\ell_1$ error loss as a natural choice to replace its $\ell_2$ counterpart, e.g., \cite{baccini19961}. However, \cite{maronna2006robust} point out that the $\ell_1$ estimate is sensitive to the bad leverage points.

% \cite{verboon1994resistant} and Although the monotone $M$-estimators can provide a unique global optimal solution, they are also incapable of handling the bad leverage points. \cite{maronna2008robust} proposed to use a redescending $\psi$ function following the $MM$ estimation method \citep{yohai1987high}. However, a robust initial estimate has to be provided, and such methods are easily trapped in local minima.

\section{Mathematical Formulation}
\label{sec:formulation_rocpca}
\subsection{Motivation and model description}% of $\bsbX$
%%%% Short version %%%%%
%The non-robustness issue of PCA has been noticed long before and  extensively investigated in robust statistics. See, for example,  \cite{maronna2006robust}\ for a comprehensive introduction. In this work, we  classify the robust PCA approaches into five classes: 1) \emph{robust covariance matrix} based methods, such as  \cite{maronna1976robust} and \cite{rousseeuw1985multivariate}; 2) \emph{projection} based methods, e.g., \cite{li1985projection} and \cite{hubert2002fast}; 3) \emph{hybrid} projection-covariance estimation based methods \citep{hubert2005robpca}; 4) \emph{spherical/elliptical} PCA \citep{locantore1999robust}; 5) \emph{low rank matrix approximation} based methods,  \cite{croux2003fitting}, \cite{Candes2011}, \cite{zhou2010stable} among others.  Due to space limitations, we give a  more detailed literature review in  the supplementary material.
%%%% Short version ends %%%

Let $\bsbX$ be an $n \times p$ data matrix with $n$ observations in $p$-dimensional space. Assume no outliers exist for now and $\bsbV^o \in \mathbb{R}^{p\times r}$ consists of the top $r$ ideal PC loading vectors. Then $\bsbP_{\bsbV^o} = \bsbV^o {\bsbV^o}^T$ defines the $r$-dimensional PC subspace. Recall the characterization of PCA via low rank matrix approximation:
$%\begin{equation}
%\label{eq:motivationFormulation}
\min_{\bsbB} \|\bsbX-\bsbB\|_F^2 \quad \text{s.t.} \quad \text{rank}(\bsbB)\le r.
$ %\end{equation}
The optimal $\bsbB$ must lie in the PC subspace, i.e., $\bsbB\bsbP_{\bsbV^o}=\bsbB$. Decomposing $\bsbX$ into $\bsbX\bsbP_{\bsbV^o}$ and $\bsbX(\bsbI-\bsbP_{\bsbV^o})$ (the projections of $\bsbX$ onto the PC subspace and  OC subspace, respectively), the objective function %in \eqref{eq:motivationFormulation}
 becomes
\begin{equation}
\label{eq:motivationDecomposition}
\|\bsbX - \bsbB\|_F^2=\|\bsbX\bsbP_{\bsbV^o}-\bsbB\|_F^2+\|\bsbX(\bsbI-\bsbP_{\bsbV^o})\|_F^2.
\end{equation}

Now suppose outliers do exist, in the situation of which the above  may result in a misleading $\bsbB$-estimate, due to the non-robust nature of the $\ell_2$ loss function.
%To robustify~\eqref{eq:motivationFormulation}, let's examine the two terms on the right hand side of~\eqref{eq:motivationDecomposition}.
The robustification of the first term $\|\bsbX\bsbP_{\bsbV^o}-\bsbB\|_F^2$ is related to the PC outliers (cf. Figure~\ref{fig:PCoutlier}), and no matter what robust loss is chosen, one can always set $\bsbB^o=(\bsbX\bsbV^o)\bsbV^{oT}$ to satisfy the low rank constraint. Thus the first term always vanishes in optimization, regardless of the choice of the loss. In contrast, the second term $\|\bsbX(\bsbI-\bsbP_{\bsbV^o})\|_F^2$ is independent of $\bsbB$, and its robustification is to address pertinent outliers in the OC subspace (cf. Figure~\ref{fig:OCoutlier}). Therefore, the crux in robust PC estimation lies in incorporating  OC outliers into the second term.

% \subsection{Model description} \label{sec:modeldescription_rocpca} Given a contaminated data matrix $\bsbX \in \mathbb{R}^{n\times p}$, with $n$ observations of dimension $p$, Motivated by the idea that we can project the data onto a proper subspace and then eliminate the anomalies in that subspace (see Figure~\ref{fig:OCoutlier}), we introduce a \emph{projected} mean-shift outlier model with unknown $\bsbVp$, $\bsbmu$, and $\bsbS$:

Motivated by this, we introduce a \emph{projected} mean-shift outlier model:
\begin{equation}
\label{eq:MStransformed}
\bsb{XV}_{\perp}^*=\bsb{1} \bsb{\mu}^{*T} +\bsb{S}^* + \bsb{E}.
\end{equation}
We use $d := p-r$ throughout this paper. $\bsbV_{\perp}^*$ is a $p\times d$ matrix satisfying $\bsbV_{\perp}^{*T}\bsbV_{\perp}^*=\bsbI$, %to characterize the subspace orthogonal to the rank-$r$ PC subspace,
and $\bsbX\bsbV_{\perp}^*$ gives the coordinates after projecting the data onto the OC subspace. In model~\eqref{eq:MStransformed}, $\bsbX\bsbV_{\perp}^*$ is decomposed into three parts: mean, outlier and noise. Concretely, 1) $\bsb{1} \bsb{\mu}^{*T}$ stands for the mean term, where $\bsb{1} = [1,1,...,1]^T \in \mathbb{R}^{n}$ and $\bsbmu^*$ is a $d$-dimensional mean vector for the transformed observations; 2) $\bsb{S}^*=[\bsb{s}_1^*,...,\bsb{s}_n^*]^T=[s_{i,j}^*] \in \mathbb{R}^{n\times d}$ is the outlier matrix, describing the outlyingness of each observation or entry; 3) finally, the noise term $\bsb{E}$ has i.i.d. $\mathcal{N}(0,\sigma^2)$  entries (or {sub-Gaussian} entries which may be dependent). The goal is to recover $\bsbmu^*$, $\bsbS^*$ and $\bsbV_{\perp}^*$ jointly. The problem for small  $r$ is seemingly more challenging, because of  the increased dimensionality of  the subspace where  `bad' OC outliers can occur.

Assume that $\bsb{S}$ is sparse (because outliers should not be the norm), then shrinkage estimation can be used: %achieved by solving the following optimization problem:
\begin{equation}
\label{eq:generalOpt}
\min_{\text{$\bsb{V}_{\perp}$, $\bsb{\mu}$, $\bsb{S}$}} \ell({\bsb{V}_{\perp},\bsb{\mu},\bsb{S}}) = \frac{1}{2} \|\bsb{XV}_{\perp}-\bsb{1} \bsb{\mu}^T-\bsb{S}\|_F^2+P(\bsb{S};\lambda) \quad \text{s.t.} \quad \bsbVpt\bsbVp=\bsbI,
\end{equation}
where $P(\bsb{S};\lambda)$ stands for  a general sparsity-promoting penalty (or constraint) with $\lambda$ as the regularization parameter. Hereinafter, the study of~\eqref{eq:generalOpt} is referred to as \emph{robust orthogonal complement principal component analysis} (\textbf{ROC-PCA}). {Different from PCP, where the sparsity is pursued in the raw observation space, ROC-PCA introduces sparsity after projecting the data points onto the OC subspace, and $\bsbS\bsbVpt$ is not necessarily sparse}. As will be shown later, ROC-PCA  provides a robust guarantee for estimating  the OC subspace (and therefore the PC subspace) and can identify the outliers simultaneously. The regularizations through rank reduction and sparsity make ROC-PCA applicable to $p\gg n$ datasets.
% Owing to the regularization vis sparsity and low rankness, ROC-PCA applies to any dataset with $p$ possibly much larger than $n$.

There are two main ways of enforcing sparsity in $\bsbS$, corresponding to element-wise (\textbf{e-Type}) outliers and row-wise (\textbf{r-Type}) outliers, respectively. The e-Type ROC-PCA is defined as
\begin{equation}
\label{eq:ewise}
 \min_{(\bsb{V}_{\perp}, \bsb{\mu}, \bsb{S})} \frac{1}{2} \|\bsb{XV}_{\perp}-\bsb{1} \bsb{\mu}^T-\bsb{S}\|_F^2+\sum_{ij}P(|s_{ij}|;\lambda_{ij}) \quad \text{s.t.} \quad \bsb{V}_{\perp}^T\bsb{V}_{\perp}=\bsb{I}.
\end{equation}
$P$ can take various forms (possibly non-convex), such as $P(\bsbS;\lambda)=\sum_{ij}\lambda_{ij} |s_{ij}|$ or $\lambda\|\bsb{S}\|_1$ when $\lambda=\lambda_{ij}$. This popular $\ell_1$ penalty~\citep{tibshirani1996regression} is however  well known to suffer from biased estimation and inconsistent selection (\cite{zou2005regularization}, \cite{zhao2006model}). Moreover, convex penalties have limited power in dealing with multiple gross outliers with high leverage values~\citep{she2011outlier}. One non-convex alternative is the $\ell_0$ penalty $P(\bsbS;\lambda)=\sum_{ij}(\lambda_{ij}^2/2)1_{s_{ij}\neq 0}$ or $(\lambda^2/2)\|\bsb{S}\|_0$ when $\lambda=\lambda_{ij}$. Some fusion penalties, such as SCAD~\citep{fan2001variable} and Hard-Ridge \citep{she2012iterative}, can also be applied. %of type `$\ell_0+\ell_2$'
%a convex choice is the basic and  namely, . Though giving the best convex approximation to the $\ell_0$ norm,
Similarly, to address outliers in a row-wise manner, we introduce the r-Type ROC-PCA
\begin{equation}
\label{eq:rwise}
 \min_{(\bsb{V}_{\perp}, \bsb{\mu}, \bsb{S})} \frac{1}{2} \|\bsb{XV}_{\perp}-\bsb{1} \bsb{\mu}^T-\bsb{S}\|_F^2+\sum_{i}P(\|\bsb{s}_i\|_2;\lambda_i) \quad \text{s.t.} \quad \bsb{V}_{\perp}^T\bsb{V}_{\perp}=\bsb{I},
\end{equation}
where $\bsb{s}_i^T$ is the $i$-th row vector of $\bsbS$. All element-wise penalties can be adapted to promote group sparsity, e.g., $\lambda \|\bsb{S}\|_{2,1}$ with $\|\bsb{S}\|_{2,1}\triangleq \sum_i \|\bsb{s}_i\|_2$, %~\citep{yuan2006model},
and $(\lambda^2/2)\|\bsb{S}\|_{2,0}$ with $\|\bsb{S}\|_{2,0}\triangleq \sum_i 1_{\bsb{s}_i \neq 0}$.
Classic robust statistics pays special attention to r-Type outliers, while  e-Type outliers may arise from a fully independent multivariate contamination model---interested readers can refer to \cite{alqallaf2009pom} for more details.

% Moreover, group SCAD, group $\ell_p$, and group hard-ridge \citep{she2012iterative} provide some grouped fusion penalties.
%  Apart from choosing
% different penalties, we are particular interested in the constrained form of ROC-PCA: $\min_{\text{$\bsb{V}_{\perp}$,
% $\bsb{\mu}$, $\bsb{S}$}} \frac{1}{2} \|\bsb{XV}_{\perp}-\bsb{1} \bsb{\mu}^T-\bsb{S}\|_F^2\text{ s.t.
% }\bsb{V}_{\perp}^T\bsb{V}_{\perp}=\bsb{I},\|\bsbS\|_0 \le q^e$, where $q^e$ directly offers an upper bound of the number of
% outliers (see Section~\ref{sec:screening_rocpca} for more details).
% Similarly, we are also interested in the corresponding constrained form of ROC-PCA, where the group sparsity penalty on $\bsbS$ is replaced with the constraint $\|\bsbS\|_{2,0} \le q$.

\subsection{ROC-PCA as generalized M-estimators}
ROC-PCA is derived by use of   the \emph{additive} robustification scheme of~\cite{she2011outlier}. The conventional way to achieve  robust estimation is through modifying the Frobenius-norm loss, or using the $M$-estimators. In this subsection, we first generalize the $M$-estimators to robust PC subspace estimation (a manifold setting), and then build a universal connection between ROC-PCA and such generalized $M$-estimators.

We begin by reviewing the definition of the $M$-estimators in linear regression $\bsby=\bsbX\bsb{\beta}+\bsb{\epsilon}$ with $\bsby=[y_1,...,y_n]^T$ and $\bsbX=[\bsbx_1,...,\bsbx_n]^T \in \mathbb R^{n\times p}$. The $\rho$-type $M$-estimator is defined to be a stationary point of $\sum_{i=1}^n \rho(y_i-\bsb{x}_i^T\bsbbeta)$, and  the (more general) $\psi$-type $M$-estimator is defined to be a solution to the equation $\bsbX^T\psi(\bsby-\bsbX\bsbbeta)=\bsb{0}$, where $\psi$, not necessarily a derivative function, is applied componentwise. %Under the assumption that $\rho$ is differentiable, one can set $\psi(t)=\rho^{\prime}(t)$.
 In our ROC-PCA setting, replacing the $\ell_2$ loss function with a  robust loss $\rho$ leads to robust PC subspace recovery:
\begin{equation}
\label{eq:rhoMestimate}
\min_{(\bsb{\mu}, \bsb{V}_{\perp})} \sum_{i=1}^n \sum_{j=1}^{d} \rho((\bsb{XV}_{\perp}-\bsb{1}\bsb{\mu}^T)_{ij};\lambda) \ \ \text{s.t.} \ \ \bsb{V}_{\perp}^T\bsb{V}_{\perp}=\bsb{I},
\end{equation}
where $\lambda$ is a parameter of the loss function. For a general $\rho$, the optimization with respect to $\bsbmu$ and $\bsbVp$ could be difficult.

A more useful $\psi$-type $M$-estimator for PC subspace recovery is defined as follows. To motivate the definition, we assume $\psi=\rho^{\prime}$, and view~\eqref{eq:rhoMestimate} as an \emph{unconstrained} optimization problem on the manifold $\Omega:=\mathbb{R}^{d}\times \mathbb{O}^{p\times d}$, where $\mathbb{R}^{d}$ denotes a $d$-dimensional Euclidean manifold, and $\mathbb{O}^{p\times d}$ represents a Stiefel manifold, the set of all $p\times d$ matrices $\bsbVp$ satisfying the orthogonality constraint $\bsbV_{\perp}^T\bsbVp=\bsbI$. The derivative of the loss with respect to $\bsbmu$ is given by $\bsb{1}^T\psi(\bsbX\bsbVp-\bsb{1}\bsbmu^T;\lambda)$. The trickier part is to define the gradient on the Stiefel manifold $\mathbb{O}^{p\times d}$. Equipped with the \emph{canonical metric} (cf. Section~\ref{sec:vpopt_rocpca}), we calculate the Riemannian gradient of $\sum_{i=1}^n \sum_{j=1}^{d} \rho((\bsb{XV}_{\perp}-\bsb{1}\bsb{\mu}^T)_{ij};\lambda)$  with respect to $\bsbVp$ (details given in the appendix):
$\bsbX^T\psi(\bsbX\bsbVp-\bsb{1}\bsbmu^T;\lambda)-\bsbVp(\psi(\bsbX\bsbVp-\bsb{1}\bsbmu^T;\lambda))^T\bsbX\bsbVp.$
Now, given any $\psi$ function, the generalized ($\psi$-type) $M$-estimator $(\hat{\bsbmu}$, $\hat{\bsbV}_{\perp})$ for ROC-PCA is defined as a solution to the following equations:%where $\psi$ is applied componentwise
\begin{equation}
\label{eq:genralMEst}
\left\{
\begin{array}{ccl}
\bsb{1}^T\psi(\bsbX\bsbVp-\bsb{1}\bsbmu^T;\lambda) & = & \bsb{0},\\
\bsbX^T\psi(\bsbX\bsbVp-\bsb{1}\bsbmu^T;\lambda)-\bsbVp (\psi(\bsbX\bsbVp-\bsb{1}\bsbmu^T;\lambda))^T\bsbX\bsbVp & = & \bsb{0}.
\end{array}
\right.
\end{equation}
Interestingly, there is a universal connection between  ROC-PCA and the generalized $M$-estimation.

\begin{theorem}
\label{mytheorem:Mestimate} (i) Let $\Theta(\cdot;\lambda)$ be an arbitrarily given \emph{thresholding rule} (cf. Section~\ref{sec:musopt_rocpca}), and  $P$  be any  penalty associated with $\Theta$ such that
\begin{equation}
\label{eq:penaltyconstruction}
%\begin{array}{l}
P(t;\lambda) - P(0;\lambda)=\int^{|t|}_0 (\sup \{s:\Theta(s;\lambda) \le u\}-u)\rd u+q(t;\lambda),
%\end{array}
\end{equation}
for some nonnegative  $q(\cdot; \lambda)$  satisfying $q(\Theta(s;\lambda);\lambda)=0$  $\forall s\in \mathbb{R}$. %\begin{equation}
%\label{eq:MandTheta}
%\Theta(t;\lambda)+\psi(t;\lambda)=t, \quad \forall t,
%\end{equation}
Suppose $(\hat{\bsbV}_{\perp}, \hat{\bsbmu}, \hat{\bsbS})$ is a {coordinate-wise minimum}  %(cf. Appendix~\ref{appendix:MThetaConnection} for the definition)
 of~\eqref{eq:ewise} and $\Theta$  is continuous at $(\bsbI-\frac{1}{n}\bsb{1}\bsb{1}^T)\bsbX\hat{\bsbV}_{\perp}+ \frac{1}{n}\bsb{1}\bsb{1}^T\hat{\bsbS}$. Then $(\hat{\bsbV}_{\perp}, \hat{\bsbmu})$ is
  necessarily a robust generalized $M$-estimator associated with $\psi$, where $\psi(t;\lambda)=t- \Theta(t;\lambda)$, $\forall t$.
  (ii) Given $0\le q < n d$,    let  $(\hat{\bsbV}_{\perp}, \hat{\bsbmu}, \hat{\bsbS})$     be any  coordinate-wise minimum of
$
%\label{eq:row_screening}
 \min_{(\bsb{V}_{\perp},\bsb{\mu},\bsb{S})} \frac{1}{2} \|\bsb{XV}_{\perp}-\bsb{1} \bsb{\mu}^T-\bsb{S}\|_F^2 \text{ s.t. }  \bsb{V}_{\perp}^T\bsb{V}_{\perp}=\bsb{I},   \|\bsb{S}\|_{0} \le q.
$ Then, after dropping $\hat \bsbS$, $(\hat{\bsbV}_{\perp}, \hat{\bsbmu})$ is also a minimizer of
$
\frac{1}{2} \sum_{k=1}^{n d-q} r_{(k)}^2     \mbox{ s.t. }   \bsb{V}_{\perp}^T\bsb{V}_{\perp}=\bsb{I}, \   \bsbR  = \bsb{XV}_{\perp}-\bsb{1} \bsb{\mu}^T,
$
where $r_{(1)}, \cdots, r_{(n d)}$ are the order statistics of the elements of $\bsbR$ satisfying $ |r_{(1)}|  \le \cdots \le |r_{(n d)}|$.
\end{theorem}

%See Appendix~\ref{appendix:MThetaConnection} for its proof.
The theorem shows the correspondence between the e-Type ROC-PCA estimator and the element-wise form of the generalized $M$-estimator~\eqref{eq:genralMEst}. This universal connection provides some guidance in choosing $P$, too. For example, redescending $\psi$-functions are recommended in robust statistics to deal with gross outliers. They correspond to non-convex penalties. The conclusion can be easily extended to the r-Type outlier case~\eqref{eq:rwise}, on the basis of \cite{she2012iterative}.

On the other hand, ROC-PCA differs from  the generalized $M$-estimation  in some significant ways. First, {without the explicit introduction of  $\bsbS$}, the generalized $M$-estimators cannot reveal  outliers  inherently. A cut-off value for the residuals (which are  not independent) has to be chosen. In contrast, based on the sparsity pattern of $\hat{\bsbS}$, ROC-PCA explicitly labels all outliers. Second, the $\lambda$ in the generalized $M$-estimation is a loss parameter, the tuning of which is usually based on large-$n$ asymptotics or worst-case studies (e.g., the breakdown point); in~\eqref{eq:generalOpt}, $\lambda$  is a regularization parameter to control the bias-variance trade-off, and is   easy to be tuned in a data-dependent manner. Third, the $\psi$ function in~\eqref{eq:genralMEst} may be non-smooth or even discontinuous, but the quadratic objective function in~\eqref{eq:generalOpt} is  smooth in $\bsbVp$. Therefore, the optimization of $\bsbVp$ in ROC-PCA can be much easier nad less computationally expensive. (For example, second-order derivative information can be possibly utilized to develop faster algorithms.) Finally, the design  of $\rho$ is most suitable and effective  in robustifying the squared error loss, while  the additive robustification, by introducing a sparse shift outlier term, naturally extends to other loss functions, such as Bernoulli, Poisson, hinge loss and others.% associated with the Generalized Linear Models.
%e.g., the Newton's method, for the $\bsbVp$ optimization in ROC-PCA.  in the $\bsbVp$-optimization of ROC-PCA,

\subsection{Estimation of PC directions }
Once  $\bsbhatVp$ is obtained, the  PC subspace estimate is given by $\hat{\bsbP}=\bsb{I}-\hat{\bsbV}_{\perp}\hat{\bsbV}_{\perp}^T$. This suffices in many PCA applications, such as data visualization. Sometimes,  one may want to obtain each individual PC direction ordered in terms of importance. Under the assumption that only pure OC outliers exist, simply applying SVD to $\bsbX\hat{\bsbP}$ completes the task.

On the other hand, if one suspects that  OC outliers and  PC outliers coexist,  a robust PCA method can be further applied to $\bsbX \hat{\bsbP}$. Here, ROC-PCA also offers some computational benefits. Indeed, because $r \ll p$ and $\bsbX \hat{\bsbP}$ is free of OC outliers, a rank-$r$ SVD reduction (Section~\ref{sec:srocpca_rocpca}) can be safely performed before running robust PCA, to reduce time and space complexity. Alternatively, one can adopt a \emph{sequential} ROC-PCA scheme to extract the most important PC directions. First, apply ROC-PCA to $\bsbX_1:=\bsbX$ with the resultant robust rank-$1$ PC subspace denoted by $\hat{\bsbP}$. A spectral decomposition on $\hat{\bsbP}$  yields $\hat{\bsbv}_1$. Then ROC-PCA can be  repeatedly applied to the deflated matrix $\bsbX_k = \bsbX_{k-1}-\bsbX_{k-1}\hat{\bsb{v}}_{k-1}\hat{\bsb{v}}_{k-1}^T$ to get the rest PC directions $\hat{\bsb{v}}_k$ $(2\le k \le r)$.
% Yet this requires ROC-PCA to run $r$ times. Yet this could be costly when $r$ is not very small,

\section{Computation}
\label{sec:computation_rocpca}
The computation of ROC-PCA defined in \eqref{eq:generalOpt} is challenging  due to the orthogonality constraint, in addition to the non-smooth and possibly non-convex $P$.
%\textbf{It is worth pointing out that the orthogonality condition increases the difficulty in both computation and theory.}
In this section, we develop an alternating optimization algorithm based on Stiefel manifold optimization and iterative nonlinear thresholdings.

\subsection{$\bsbVp$-optimization}
\label{sec:vpopt_rocpca}
Given $\bsb{\mu}$ and $\bsb{S}$, minimizing $\ell$ (cf.~\eqref{eq:generalOpt}) with respect to $\bsbVp$ reduces to:
\begin{equation}
\min_{\bsbVp} f(\bsbVp) = \frac{1}{2}\|\bsb{XV}_{\perp}-\bsb{1} \bsb{\mu}^T-\bsb{S}\|_F^2 \quad \text{s.t.} \quad \bsbVpt\bsbVp=\bsb{I}.
\label{eq:StiefelOpt}
\end{equation}%notice that ,
There are many ways of solving the problem.
Our goal is to design a fast  algorithm even in  high dimensions.

Instead of treating~\eqref{eq:StiefelOpt} as a constrained optimization problem by introducing a few Lagrangian multipliers, we view it as an \emph{unconstrained} optimization problem on the Stiefel manifold $\mathbb{O}^{p\times d}:=\{\bsbVp\in \mathbb{R}^{p\times d}:\bsbVpt\bsbVp=\bsb{I}\}$, to take advantage of the smoothness of $f$ in $\bsbVp$. Optimization on the Stiefel manifold requires preserving the orthogonality constraint in updating $\bsbVp$. Our updating scheme is based on the idea of \emph{retraction}, which smoothly maps the tangent space $\mathcal{T}_{\bsbVp}(\mathbb{O}^{p\times d}):=\{\bsbDelta \in \mathbb{R}^{p\times d}:\bsbVpt\bsbDelta+\bsbDelta^T\bsbVp=0\}$ ($\mathcal{T}_{\bsbVp}$ for notational simplicity) onto the Stiefel manifold $\mathbb{O}^{p\times d}$, see, e.g.,~\cite{absil2008optimization}.

We begin by defining a Riemannian gradient of $f$ with respect to $\bsbVp$, denoted by $\nabla f$. Following~\cite{edelman1998geometry}, we adopt the \emph{canonical metric} $g_c(\bsbDelta,\bsbDelta):=\tr(\bsbDelta^T(\bsbI-\frac{1}{2}\bsbVp\bsbVpt)\bsbDelta)$. The Riemannian gradient $\nabla f$ is then defined as the unique element in $\mathcal{T}_{\bsbVp}$ such that $g_c(\nabla f, \bsbDelta) = \tr(\bsbG^T \bsbDelta)$ for any $\bsbDelta \in \mathcal{T}_{\bsbVp}$, where $\bsbG$ denotes the Euclidean gradient of $f$ with respect to $\bsbVp$, i.e., $\bsbG_{ij}=\frac{\partial{f(\bsbVp)}}{\partial{\bsbVp}_{ij}}$. It is not difficult to show that
\begin{equation}
\label{eq:Rgradient}
 \nabla f = \bsbW\bsbVp, \text{ with } \bsbW=\bsbG\bsbVpt-\bsbVp\bsbG^T \text{, } \bsbG = \bsbX^T(\bsb{XV}_{\perp}-\bsb{1}\bsb{\mu}^T-\bsb{S}).
\end{equation}

A valid updating scheme should guarantee that the new trial point lies on the manifold. Let $\bsbVp(\tau)$ be a function determining the new trial point with $\tau$ as the step size.  We use a   \textbf{Cayley transformation} based update due to  \cite{wen2010feasible}:
\begin{equation}
\label{eq:cayley}
\bsbVp(\tau)=(\bsbI+\frac{\tau}{2}\bsbW)^{-1}(\bsb{I}-\frac{\tau}{2}\bsb{W})\bsbVp.
\end{equation}
It can be verified that the curve generated by~\eqref{eq:cayley} always lies on the manifold for any $\tau$, and $\bsbVp(\tau)$ is a descent curve passing the point $\bsbVp(0)=\bsbVp$. Yet the inversion of the $p\times p$ matrix $(\bsbI+\frac{\tau}{2}\bsbW)$ in~\eqref{eq:cayley} may be  expensive when $p$ is large. When $d<p/2$, one can write $\bsbW=\bsbA_1\bsbA_2^T$ with $\bsbA_1=[\bsbG,  \bsbVp]$ and $\bsbA_2=[\bsbVp,  -\bsbG]$, and apply the matrix inversion formula to get $\bsbVp(\tau) = \bsbVp - \tau  \bsbA_1 (\bsbI + \tau \bsbA_2^T \bsbA_1/2)^{-1} \bsbA_2^T\bsbVp$ (cf. Lemma 4,~\cite{wen2010feasible}). This fast-update formula involves the inversion of a $2d\times 2d$ matrix, and turns out to be pretty useful in the design of batch ROC-PCA in Section~\ref{sec:srocpca_rocpca}.  In the case of $d\ge p/2$, one possible idea is to approximate $\bsbW$ by the product of two low rank matrices (Lemma 5,~\cite{wen2010feasible}).
% we can still apply the matrix inversion formula to obtain a descent curve  $\bsbVp(\tau)$
%Sherman-Morrison-Woodbury (SMW) formula
% By SMW, the inverse of the $p\times p$ matrix $(\bsbI+\frac{\tau}{2}\bsbW)$ can be obtained by calculating , and leads
% to a closed form updating scheme:
% \begin{equation}
% \label{eq:SMV}
% \end{equation}
% . This quasi-geodesic update is much cheaper per
% iteration than the popular schemes based on  matrix exponentiation \citep{edelman1998geometry,abrudan2008steepest} or
% polar decomposition \citep{absil2008optimization}, especially when $d<p/2$.

It remains to specify a proper step size $\tau$ to guarantee the convergence and efficiency in large  problems. We use a \textit{nonmonotone} line search scheme together with Barzilai-Borwein stepsize (BB) (\cite{barzilai1988two} and \cite{raydan1997barzilai}). In comparison with other commonly used inexact line searches, BB does {not} guarantee descent in function value at each step, but results in quick convergence and performs well in large-scale nonlinear optimization (\cite{zhang2004nonmonotone}, \cite{dai2005projected}, \cite{zhou2006gradient}). In addition, the nonmonotone search scheme only performs backtracking occasionally, and thus saves a lot of computational time. (Be aware  that cost of generating a trial point on the manifold is not cheap.)

% than inexact line searches

% Furthermore, in our problem, it is relatively expensive to generate trial points on the manifold. Rather than performing line search each time to find an acceptable stepsize, as in other inexact line searches, the nonmonotone line search only require to do so occasionally and the computation of BB stepsizes is very efficient. Therefore, the BB-based method can greatly reduce the computational cost.

% for any dimensional strictly convex quadratics~\citep{}
% other variants see, e.g.,~\cite{dai2005projected} \cite{zhang2004nonmonotone} \cite{dai2001adaptive}

% approximating the secant equation by
In more detail, the BB calculation at the $k$-th iteration requires solving $\min_{\tau_k} \|\tau_k^{-1} \delta_k(\bsbVp)-\delta_k(\nabla f)\|_F^2$ and $\min_{\tau_k} \|\delta_k(\bsbVp)-\tau_k \delta_k(\nabla f)\|_F^2$, with $\delta_k(\bsbVp)=\bsbV_{\perp}^{(k)}-\bsbV_{\perp}^{(k-1)}$ and $\delta_k(\nabla f)=\nabla f(\bsbV_{\perp}^{(k)})-\nabla f(\bsbV_{\perp}^{(k-1)})$. This leads to $\tau_k^0 = \frac{\tr(\delta_k(\bsbVp)^T\delta_k(\bsbVp))}{|\tr(\delta_k(\bsbVp)^T\delta_k(\nabla f))|}$ and $\tau_k^1 = \frac{|\tr(\delta_k(\bsbVp)^T\delta_k(\nabla f))|}{\tr(\delta_k(\nabla f)^T\delta_k(\nabla f))}$, respectively. The two solutions are used alternatively in odd and even numbered iterations. Because of the nonmonotonic behavior of BB, Raydan's adaptive nonmonotone  search scheme is applied to ensure global convergence. That is, compute the stepsize $\tau^{(k)}=\kappa^{m_k}\tau_k^i$ ($i=0$ for even $k$ and $i=1$ otherwise), where $\kappa \in (0,1)$ and $m_k$ is the smallest integer satisfying
% such that the adaptive nonmonotone line search condition
\begin{equation}
% f(\bsbV_{\perp}^{(k)}(\tau_k)) \le f_r +\rho \tau_k f^{\prime}_{\tau}(\bsbV_{\perp}^{(k)}(0))
f(\bsbV_{\perp}^{(k)}(\tau^{(k)}) \le \max_{0\le j \le \min(k,T)} f(\bsbV_{\perp}^{(k-j)}) +\rho \tau_k f^{\prime}(\bsbV_{\perp}^{(k)}(0)).
\label{eq:armijo}
\end{equation}
This  criterion uses $T$ most recent function values. It is easy to get $f^{\prime}(\bsbV_{\perp}(0)):=\frac{\partial f(\bsbVp(\tau))}{\partial \tau}|_{\tau=0}=\tr\{(\frac{\partial f(\bsbVp(\tau))}{\partial \bsbVp_{ij}(\tau) })^T (\frac{\partial \bsbVp(\tau)}{\partial \tau})\}|_{\tau=0}=-\tr\{\bsbG^T(\bsbG\bsbV_{\perp}^T-\bsbVp\bsbG^T)\bsbVp\}=-\frac{1}{2}\|\bsbW\|_F^2$, where $\bsbW$ is calculated according to~\eqref{eq:Rgradient}. In practice, we recommend $\kappa=0.1$, $T=10$ and $\rho=1e-3$. %The overall $\bsbVp$-optimization is detailed in Step 2 of Algorithm~\ref{alg:ROCPCAAlgorithm}.

How to choose the starting point  is important. Our  initialization of $\bsbV_{\perp}^{(0)}$ uses  the multi-start strategy  by \cite{rousseeuw1999fast}. First generate $m_0$ candidate $\bsbV_{\perp}^{(0)}$ at random; starting with each, run the computational algorithm %Algorithm~\ref{alg:ROCPCAAlgorithm}
 for $n_0$ iterations;  pick the best $m_1$ candidates (evaluated by the cost function value) and continue the algorithm till convergence. The final estimate $\hat{\bsbV}_{\perp}$ is the one that delivers the minimal cost function value. For example, in implementation of r-Type ROC-PCA, we use $m_0=10$, $n_0=2$ and $m_1=2$.

\subsection{$(\bsbmu,\bsbS)$-optimization}
\label{sec:musopt_rocpca}
Fixing $\bsb{V}_{\perp}$, \eqref{eq:generalOpt} reduces to
$%\begin{equation}
%\label{eq:thetaEstimate}
 \min_{\bsbmu,\bsb{S}} g(\bsbmu,\bsbS) = \frac{1}{2} \|\bsb{XV}_{\perp}-\bsb{1} \bsb{\mu}^T-\bsb{S}\|_F^2+P(\bsb{S};\lambda).
$ %\end{equation}
The optimization for $\bsbmu$ is an OLS problem with the solution given by $\bsbmu^o=\frac{1}{n}(\bsb{XV}_{\perp}-\bsb{S})^T\bsb{1}$. The $\bsbS$-optimization involves sparsity-inducing penalties (element-wise or row-wise) which are non-differentiable and possibly non-convex (corresponding to redescending $\psi$-type $M$-estimators).
% In particular, to reject gross outliers, statisticians usually favor non-convex $P$.

To give a  general algorithmic framework, we solve the problem from the viewpoint of thresholding rules. A \textbf{thresholding rule}, denoted by $\Theta$, is defined to be an \emph{odd monotone unbounded shrinkage} function \citep{she2009thresholding}. Given any $\Theta$, its vector or matrix version (still denoted by $\Theta$) is defined componentwise. For any $\bsb{s} \in \mathbb{R}^d$, the \emph{multivariate} version of $\Theta$, denoted by $\vec{\Theta}(\bsb{s};\lambda)$, is defined to be $\frac{\bsb{s}}{\|\bsb{s}\|_2}\Theta(\|\bsb{s}\|_2;\lambda)$ if $\bsb{s} \neq \bsb{0}$ and otherwise $\bsb{0}$ (cf.  \cite{she2012iterative}). For any $\bsb{S} = [\bsb{s}_1,...,\bsb{s}_n]^T \in \mathbb{R}^{n\times d}$, $\vec{\Theta}(\bsb{S};\lambda)=[\vec{\Theta}(\bsb{s}_1;\lambda),...,\vec{\Theta}(\bsb{s}_n;\lambda)]^T$.

Given an arbitrary thresholding rule $\Theta$, and let $P$ be any function satisfying~\eqref{eq:penaltyconstruction} with $q(t;\lambda)$ nonnegative and $q(\Theta(s;\lambda);\lambda)=0$ for all $s\in \mathbb{R}$. Then the minimization problem
$\min_{\bsbS} \frac{1}{2} \|\bsbY-\bsbS\|_F^2+\sum _{i}P(\|\bsb{s}_{i}\|_2;\lambda)$
has a globally optimal solution  ${\bsbS}^o =\vec{\Theta}(\bsbY;\lambda)$. Similarly, a globally optimal solution to $\min_{\bsbS} \frac{1}{2} \|\bsbY-\bsbS\|_F^2+\sum _{ij}P(|s_{ij}|;\lambda)$ is ${\bsbS}^o=\Theta(\bsbY;\lambda)$. See  Lemma 1 in~\cite{she2012iterative} for a justification (where the `continuity assumption' of $\Theta$ at $\bsbY$ is not needed because we do not require the uniqueness of $\bsbS^o$). Starting with various thresholding rules, \eqref{eq:penaltyconstruction} covers all commonly used penalties, including $\ell_1$, $\ell_0$,  $\ell_p$ ($0<p<1$),  `$\ell_0+\ell_2$'~\citep{she2012iterative} and so on. Based on such $\Theta$-$P$ coupling, a general iterative algorithm for updating $\bsbmu$ and $\bsbS$ can be designed, which is illustrated below for the r-Type problem:  %The  e-Type algorithm simply uses $\Theta$ in place of $\vec{\Theta}$.

%\mbox{}
\hspace{2.1cm}\textbf{repeat}

\hspace{2cm} $\quad \quad \bsbmu^{(k)} \leftarrow \frac{1}{n}(\bsbX \bsbVp-\bsbS^{(k)})^T\bsb{1}$

\hspace{2cm} $\quad \quad \bsbS^{(k+1)} \leftarrow \vec{\Theta}(\bsbX\bsbVp-\bsb{1} (\bsbmu^{(k)})^T;\lambda)$

\hspace{2cm} $\quad \quad k \leftarrow k+1$

\hspace{2.1cm}\textbf{until} $\|\bsbS^{(k)}-\bsbS^{(k-1)}\|$ is small enough %\\ %<\varepsilon
%\mbox{}

Clearly,   $\bsbmu^{(k)}$ does not have be   explicitly calculated:
% Moreover, one can just update $\bsbS$ during the iterations by
$
\bsbS^{(k+1)} \leftarrow \vec{\Theta}((\bsbI-\frac{1}{n}\bsb{1}\bsb{1}^T)\bsbX\bsbVp+\frac{1}{n}\bsb{1}\bsb{1}^T\bsbS^{(k)};\lambda).
$
% until convergence, and update $\bsbmu$ at the end.\\
%\vskip 0.25cm
A summary of the  complete algorithm for the r-Type ROC-PCA is shown in
%%%%% Short version %%%%
%the supplementary material. Simply replacing $\vec{\Theta}$ with its componentwise version $\Theta$ gives the e-Type ROC-PCA algorithm.
%%%% Short version ends %%%%
%%% Long version %%%%%
 Algorithm \ref{alg:ROCPCAAlgorithm}.
This alternating optimization guarantees the function-value decreasing property at each step, $g(\bsbmu^{(k)},\bsbS^{(k)}) \ge g(\bsbmu^{(k)},\bsbS^{(k+1)}) \ge g(\bsbmu^{(k+1)},\bsbS^{(k+1)})$  for any $k \ge 0$.
 Simply replacing $\vec{\Theta}$ with its componentwise version $\Theta$ solves the e-Type ROC-PCA.
% \IncMargin{1em}
% \linesnumbered
\restylealgo{ruled}
\begin{algorithm}
    \label{alg:ROCPCAAlgorithm}

    \dontprintsemicolon
    \SetKwInOut{Input}{Input}
    \SetKwInOut{Output}{Output}
    \caption{The (r-Type) ROC-PCA Algorithm}

    \Indentp{-0.5em}
    \Input{Data matrix $\bsbX$; thresholding rule $\Theta(\cdot;\lambda)$; initial estimate $\bsbV_{\perp}^{(0)}$ (cf. Section~\ref{sec:vpopt_rocpca});  nonmonotone line search parameters $\rho$, $\kappa$, $T$ (cf. Section~\ref{sec:vpopt_rocpca}).}
    \Indentp{0.5em} %maximum numbers of outer and inner iterations $M_{outer}$, $M_{inner,1}$, $M_{inner,2}$; error tolerances $\varepsilon_{outer}$, $\varepsilon_{inner,1}$, $\varepsilon_{inner,2}$, $\varepsilon_{inner,3}$;
        \BlankLine
        \emph{Initialization:}\\%$\mathfrak{0}$
        $i \leftarrow 0$, $\bsbV_{\perp}^{(i)} \leftarrow \bsbV_{\perp}^{(0)}$, $\bsbS^{(i)} \leftarrow \bsb{0}$\\

        % \Repeat{$\|\bsbV_{\perp}^{(i)}(\bsbV_{\perp}^{(i)})^T-\bsbV_{\perp}^{(i-1)}(\bsbV_{\perp}^{(i-1)})^T\|_{\infty}/p< \varepsilon_{outer}$  or $i > M_{outer}$}{
        \Repeat{$\|\bsbV_{\perp}^{(i)}(\bsbV_{\perp}^{(i)})^T-\bsbV_{\perp}^{(i-1)}(\bsbV_{\perp}^{(i-1)})^T\|_{\infty}/p$ \text{\upshape is small or} $i$ \text{\upshape is large enough}}{
            \BlankLine
            \emph{$\mathfrak{1}$: $(\bsbmu,\bsbS)$-optimization via $\Theta$-estimation}\\
            $k \leftarrow 0$, $\bar{\bsb{S}}^{(k)} \leftarrow \bsb{S}^{(i)}$\\

            % \emph{$\mathfrak{2.1}$: Optimize $\bsbS$}\\
            % \Repeat{$\|\bar{\bsbS}^{(k)}-\bar{\bsbS}^{(k-1)}\|_{\infty}< \varepsilon_{inner,1}$ or $k > M_{inner,1}$}{
            \Repeat{$\|\bar{\bsbS}^{(k)}-\bar{\bsbS}^{(k-1)}\|_{\infty}$ \text{\upshape is small or} $k$ \text{\upshape is large enough}}{
                $\bar{\bsbS}^{(k+1)} \leftarrow \vec{\Theta}((\bsbI-\frac{1}{n}\bsb{11}^T)\bsbX\bsbV_{\perp}^{(i)}+\frac{1}{n}\bsb{11}^T\bar{\bsbS}^{(k)}); \lambda)$ \label{lineno_thetaEst}\\
                $k \leftarrow k+1$\\
            }
            $\bsb{S}^{(i+1)} \leftarrow \bar{\bsb{S}}^{(k)}$, $\bsb{\mu}^{(i+1)} \leftarrow \frac{1}{n}(\bsbX\bsbV_{\perp}^{(i)}-\bsb{S}^{(i+1)})^T \bsb{1}$\\
            % \emph{$\mathfrak{2.1}$: Update $\bsbmu$}\\
            % $\bsb{\mu}^{(i+1)} \leftarrow \frac{1}{n}(\bsbX\bsbV_{\perp}^{(i)}-\bsb{S}^{(i+1)})^T \bsb{1}$\\

            \BlankLine
            \BlankLine
            \emph{$\mathfrak{2}$: $\bsbVp$-optimization with alternating BB stepsizes and non-monotone line search }\\
            % $f_r \leftarrow +\infty$, $f_* \leftarrow f(\bsbV_{\perp}^{(i)})$, $f_c \leftarrow f(\bsbV_{\perp}^{(i)})$ \\
            $k \leftarrow 0$, $\bar{\bsbV}^{(k)}_{\perp} \leftarrow \bsbV_{\perp}^{(i)}$, $\bsbJ \leftarrow \bsb{1}(\bsbmu^{(i+1)})^{T}+\bsbS^{(i+1)}$, $f^{(0)}\leftarrow \|\bsbX\bar{\bsbV}^{(0)}_{\perp}-\bsbJ\|_F^2$\\%$\bsbG \leftarrow \bsbX^T(\bsbX\bsbV_{\perp}^{(i)}-\bsb{1}(\bsbmu^{(i)})^{T}-\bsbS^{(i)})$
                % \Repeat{$\|\nabla f\|_F< \varepsilon_{inner,2}$ or $\frac{|f^{(k)}-f^{(k-1)}|}{|f^{(k-1)}|}< \varepsilon_{inner,3}$ or $k>M_{inner,2}$}{
                \Repeat{$\|\nabla f\|_F$ \text{\upshape or} $|f^{(k)}-f^{(k-1)}| / |f^{(k-1)}|$ \text{\upshape is small or} $k$ \text{\upshape is large enough}}{
                    \BlankLine
                    \emph{$\mathfrak{2.}$a): BB-update}\\
                    % \lIf{$k=0$}{$\tau \leftarrow 0.5$} \lElse{for even $k$, $\tau \leftarrow \tau_k^0$; otherwise , $\tau \leftarrow \tau_k^1$}
                    If $k=0$, $\tau \leftarrow 0.5$, else $\tau \leftarrow \tau_k^0$ for even $k$, otherwise $\tau \leftarrow \tau_k^1$   (see Section~\ref{sec:vpopt_rocpca})\\
                    % If $k$ is even, $\tau \leftarrow \tau_k^0$ ; otherwise, $\tau \leftarrow \tau_k^1$ (see Section~\ref{sec:vpopt_rocpca} for $\tau_k^0$ and $\tau_k^1$)\\

                    \BlankLine
                    \emph{$\mathfrak{2.}$b): Backtracking} \\
                    $\bsbG \leftarrow \bsbX^T(\bsbX\bar{\bsbV}^{(k)}_{\perp}-\bsbJ)$, $\nabla f \leftarrow \bsbG - \bar{\bsbV}^{(k)}_{\perp} \bsbG^T \bar{\bsbV}^{(k)}_{\perp}$ \\
                    $f_r \leftarrow \max_{0\le j \le \min(k,T)} \{f^{(k-j)}\}$, $f^{\prime} \leftarrow -\frac{1}{2}\|\bsbG(\bar{\bsbV}^{(k)}_{\perp})^T-\bar{\bsbV}^{(k)}_{\perp}\bsbG^T\|_F^2$\\
                    \Repeat{$f^{(k+1)} < f_r +\rho \tau f^{\prime}$}{%f^{\prime}_{\tau}(\bsbV_{\perp}^{(i)})
                        $\widetilde{\bsbV}_{\perp} \leftarrow \bar{\bsbV}^{(k)}_{\perp} - \tau  \bsbA_1 (\bsbI + \tau \bsbA_2^T \bsbA_1/2)^{-1} \bsbA_2^T\bar{\bsbV}^{(k)}_{\perp}$, where $\bsbA_1=[\bsbG$, $\bar{\bsbV}^{(k)}_{\perp}]$ and $\bsbA_2=[\bar{\bsbV}^{(k)}_{\perp}$, $-\bsbG]$\\
                        % $f^{(k+1)} \leftarrow \|\bsbX\widetilde{\bsbV}_{\perp}-\bsb{1}(\bsb{\mu}^{(i+1)})^{T}-\bsb{S}^{(i+1)}\|_F^2$ \\
                        $f^{(k+1)} \leftarrow \|\bsbX\widetilde{\bsbV}_{\perp}-\bsbJ\|_F^2$ \\
                        $\tau \leftarrow \kappa \tau$
                    }
                    % Update $f_r$ according to Section~\ref{sec:vpopt_rocpca}\\%Algorithm~\ref{alg:RefValUpdate}\\
                    $\bar{\bsbV}^{(k+1)}_{\perp} \leftarrow \widetilde{\bsbV}_{\perp}$\\

                %\STATE If $f^{(k+1)}<f_*$, set $f_* \leftarrow f^{(k+1)}$, $f_c\leftarrow f^{(k+1)}$, $t \leftarrow 0$;
               % otherwise, set $f_c \leftarrow \max(f_c,f^{(k+1)})$, $t \leftarrow t+1$, and if $t=T$, set $f_r \leftarrow f_c$, $f_c \leftarrow f^{(k+1)}$, $t \leftarrow 0$

                    $k \leftarrow k+1$\\
                }

            $\bsbV_{\perp}^{(i+1)} \leftarrow \bar{\bsbV}^{(k)}_{\perp}$\\
            $i \leftarrow i+1$\\
            \BlankLine

        }
    \BlankLine
    \Indentp{-0.5em}
    \Output{$\hat{\bsb{V}}_{\perp}=\bsbV_{\perp}^{(i)}$, $\hat{\bsb{S}}=\bsb{S}^{(i)}$ and $\hat{\bsb{\mu}}=\bsb{\mu}^{(i)}$}

\end{algorithm}
% \DecMargin{1.5em}
%%% Long version ends %%%%%

\paragraph{Progressive quantile thresholding based iterative screening}
%\label{sec:screening_rocpca} %the penalized ROC-PCA
The penalty parameter $\lambda$ in~\eqref{eq:generalOpt} adjusts the bias-variance trade-off, but is inconvenient if one wants to specify its value directly. Here, we propose some constrained forms of ROC-PCA to address the issue. For the r-Type outliers, consider
\begin{equation}
 \min_{(\bsb{V}_{\perp},\bsb{\mu},\bsb{S})} \frac{1}{2} \|\bsb{XV}_{\perp}-\bsb{1} \bsb{\mu}^T-\bsb{S}\|_F^2+\frac{\eta}{2}\|\bsb{S}\|_F^2\quad \text{s.t.} \quad \bsb{V}_{\perp}^T\bsb{V}_{\perp}=\bsb{I}, \  \|\bsb{S}\|_{2,0} \le q,\label{eq:row_screening}
\end{equation}
where, in addition to the ridge penalty to account for large noise and clustered outliers (collinearity), the group $\ell_0$ \textbf{constraint} is imposed on $\bsb{S}$ rather than a penalty. Similarly,  $\|\bsb{S}\|_0 \le q^e$, gives the constrained form of the e-Type ROC-PCA. They extend the LTS \citep{Roussbook}  due to Part (ii) of Theorem \ref{mytheorem:Mestimate}.  {Unless otherwise specified, we use the {constrained} forms of ROC-PCA in  computer experiments.} Compared with  the penalty parameter $\lambda$ in  \eqref{eq:rwise} or \eqref{eq:ewise}, $q$ (or $q^e$), as an upper bound of the number of outliers, is both meaningful and intuitive in robust analysis. Nicely,  $q$ is not a sensitive parameter to subspace recovery, as long as it is within a reasonable range (see Section \ref{sec:simmulationQtuning_rocpca}). %Section~\ref{sec:simmulationQtuning_rocpca}).
The ridge shrinkage parameter $\eta$ is even more insensitive and its search  grid can be small---in implementation, we simply fix $\eta$ at a small value, say, 1e-3.

The constrained ROC-PCA shares the same $\bsbVp$-optimization with the penalized form. As for the  $\bsbS$-optimization, fortunately, we can adapt the $\Theta$-estimators  to this subproblem via a \emph{quantile} thresholding rule $\Theta^{\#}(\cdot;q^{e},\eta)$.
%We use the r-Type as an instance to describe the algorithm  for solving the constrained optimization; the  discussion applies to the e-Type as well.
 For any $\bsbS=[\bsbs_1,\cdots,\bsbs_n] \in \mathbb{R}^{n\times d}$ and $1 \le q^{e} \le nd$, $\Theta^{\#}(\bsbS;q^{e},\eta)$ shrinks the top $q^{e}$ largest entries (in absolute value) of $\bsbS$ by a factor of $1+\eta$, and sets all remaining entries to be $0$. A \emph{multivariate} version $\vec{\Theta}^{\#}$ to be used for the constrained r-Type ROC-PCA problem is defined as $\vec{\Theta}^{\#}(\bsb{S};q,\eta)=\mbox{diag}\{\Theta^{\#}(g(\bsb{S});q,\eta)\}\bsb{S}^o$, where $g(\bsb{S})=[\|\bsb{s}_i\|_2]_{n\times 1}$ and $\bsb{S}^o=(\mbox{diag}\{g(\bsb{S})\})^+\bsb{S}$ with $^+$ standing for the Moore-Penrose pseudoinverse. Now, for the r-Type problem, we can use  $\vec{\Theta}^{\#}$ in place of $\vec{\Theta}$ and run
$%\begin{equation}
%\label{eq:quantileUpdate}
\bar{\bsbS}^{(k+1)} \leftarrow \vec{\Theta}^{\#}((\bsbI-\frac{1}{n}\bsb{11}^T)\bsbX\bsbV+\frac{1}{n}\bsb{11}^T\bar{\bsbS}^{(k)}; q,\eta).
$ %\end{equation}
The $\bsbS$-update not only  guarantees the  non-increasing of the function value but satisfies $\|\bsbS\|_{2, 0} \le q$  \citep{SheSpec}. To lessen greediness, we advocate  \textit{progressive} quantile-thresholding based iterative screening. Define a monotone sequence of integers $\{q(k)\}$ decreasing from $n$ to the target value $q$. At the $k$-th iteration, the above quantile parameter $q$ is now replaced by $q(k)$.  Empirically, $q(k) = \max (q,2n/(1+e^{\nu k}))$ with $\nu = 0.05$ gives a fast and accurate cooling scheme.

\section{Non-asymptotic Analysis of  ROC-PCA }
\label{sec:theory_rocpca}
%In this section, we theoretically evaluate the robustness and the finite sample behavior of ROC-PCA.

The finite-sample performance of ROC-PCA   is of great theoretical interest.
Due to the equivalence established in Theorem \ref{mytheorem:Mestimate}, it is not difficult for one to show some asymptotics under the classic setting where $n\rightarrow \infty$ and $r, p$ are fixed, as well as the (non-stochastic)  breakdown point properties of ROC-PCA. Nevertheless, we wish to perform  large-$p$ or even non-asymptotic robust analysis to meet the challenge of modern statistical applications.  Our tool for such  theoretical studies   is the  \textit{oracle inequalities} \citep{Donoho94}.
We take a  predictive learning perspective and study the data approximation power of ROC-PCA.
Let the model be $\bsbX = \bsbA^* {\bsbV^*}^T + \bsbS^* {\bsbV_{\perp}^*}^T + \bsbE$, where $\bsbA^* \in \mathbb R^{n \times r}$, $\bsbS^* \in \mathbb R^{n \times d}$ with $d=p-r$, $[\bsbV^*, \bsbV_{\perp}^*] \in \mathbb O^{p\times p}$. Assume all entries of $\bsbE$ are iid Gaussian $\sim \mathcal N(0, \sigma^2)$ (or {sub-Gaussian}, as in the appendix).  In this section, we ignore the intercept term for simplicity and suppose the outlier matrix $\bsbS^*$  is  row-wise sparse. The problem of interest can be formulated by
%(for simplicity,  the $\ell_2$ penalty is dropped):
\begin{align}
 \min_{\bsbS, \bsbV_\perp, \bsbA, \bsbV }\| \bsbX - \bsbA\bsbV^T - \bsbS \bsbV_{\perp}^T \|_F^2 \mbox{ s.t. }  \|\bsbS\|_{2,0}\leq q, [\bsbV, \bsbV_\perp] \in \mathbb O^{p\times p}.  \label{fullprob}
\end{align}
This is a rephrasing  of ROC-PCA.  Indeed,  the  loss can  be written in a separable form
$ \|\bsbX\bsbV  - \bsbA\|_F^2+  \|\bsbX \bsbV_{\perp}  -\bsbS \|_F^2$
 and so the optimization with respect to $\bsbS$ and $\bsbV_{\perp}$ corresponds to \eqref{eq:row_screening}.   On the other hand,  with $(\hat \bsbS, \hat \bsbV_\perp)$ available, the  optimal $\hat\bsbA  \hat\bsbV=\bsbX(\bsbI-\hat\bsbV_{\perp}\hat\bsbV_{\perp}^T)$ can be obtained afterwards.

  Given any $(\hat\bsbA, \hat\bsbS, \hat\bsbV, \hat\bsbV_{\perp})$, define its \textit{mean} approximation error   by
$$
M(\hat \bsbA, \hat\bsbS, \hat\bsbV, \hat\bsbV_{\perp};\bsbA^*, \bsbS^*, \bsbV^*, \bsbV_{\perp}^*) = \frac{1}{np}\| \hat\bsbA\hat\bsbV^T + \hat \bsbS \hat\bsbV_{\perp}^T - \bsbA^*{\bsbV}^{ * T} - \bsbS^* {\bsbV_{\perp}^{* T}}\|_F^2.
$$
or $M(\hat \bsbA, \hat\bsbS, \hat\bsbV, \hat\bsbV_{\perp})$ when there is no ambiguity.
The  approximation error  is always meaningful in evaluating  the performance of an estimator, without requiring any signal strength assumption. % Note that although  $\bsbA, \bsbV$  may  have some ambiguity issues, ,

%We  perform non-asymptotic oracle studies  for the globally optimal solutions of ROC-PCA.
\begin{theorem}\label{mytheorem:oracleerr}
Let $(\hat \bsbA, \hat\bsbS, \hat\bsbV, \hat\bsbV_{\perp})$ be any  globally optimal point of  \eqref{fullprob}. Then,   the following oracle inequality holds for any $(\bsbA, \bsbS, \bsbV, \bsbV_{\perp})$
 satisfying  $\|\bsbS\|_{2,0}\leq q$, $\bsbA\in \mathbb R^{n\times r}$, $[\bsbV, \bsbV_\perp] \in \mathbb O^{p\times p}$:
\begin{align}
% \begin{split}
          \EE\left[ M(\hat \bsbA, \hat\bsbS, \hat\bsbV, \hat\bsbV_{\perp})\right]
 \lesssim  M( \bsbA, \bsbS, \bsbV, \bsbV_{\perp}) + P_o(q, r) +\frac{ \sigma^2}{n p},
%\end{split}
\label{oracleineq}
\end{align}
where  $P_o(q, r)$ is short for $
P_o(q, r;  n, p, \sigma^2)={ \sigma^2} \{ q p + r n + r p +   q\log(en/q) \}/({n p})$,
and   $\lesssim$  denotes  an inequality that holds up to a multiplicative numerical constant.

\end{theorem}

The theorem applies to $\bsbE$ with $\vect(\bsbE)$ being sub-Gaussian (which includes  bounded random matrices, and allows for column and/or row dependencies).  \eqref{oracleineq} is in expectation form; a high probability result with the same error rate   $P_o$  can be obtained as well (without the last  additive term $\sigma^2/(np)$). See the appendix for proof details.
%%%% Long version %%%%%%
%  Appendix~\ref{appendix:oracleerr} for the proof details.
%%%% Long version ends %%%%
%Remarks.
%\begin{enumerate}
%\item
The result is  non-asymptotic  in nature and applies to any $r, q, n, p$. Note that it  does not require any incoherence condition  which is commonly assumed in the literature.
%$(\bsbA^*, \bsbS^*, \bsbV^*, \bsbV_{\perp}^*)$.
%\item

According to \eqref{oracleineq}, a sharp  risk upper bound is obtained  by taking the  infimum  of the right hand side over the set of \textit{all} valid reference signals $(\bsbA, \bsbS, \bsbV, \bsbVp)$. First, with  $\bsbS = \bsbS^*$, $\bsbA = \bsbA^*$, $\bsbV=\bsbV^*$,  such that the first term $M(\cdot)$ vanishes,  we can get  an error rate of order  ${ \sigma^2} \{ q^* p + r^* n + r^* p +   q^*\log(en/q^*) \}/(np)$ (which is optimal in a minimax sense). But  our conclusion holds more generally---in particular,     $\bsbS^*$  does not  have to be exactly  sparse.
 Indeed, when $\bsbS^*$ contains  many  small but nonzero entries,   a reference   $\bsbS$ with much reduced support  can benefit from the bias-variance trade-off to attain a lower bound than    simply taking $\bsbS =\bsbS^*$. In other words, the obtained oracle inequality ensures the ability of ROC-PCA  in dealing with  mild outliers, which is of great practical interest.

A by-product is  the  finite-sample breakdown property. First,  define the finite-sample breakdown point for an arbitrary estimator $(\hat \bsbA, \hat\bsbS, \hat\bsbV, \hat\bsbV_{\perp})$ in terms of its \textbf{risk}: Given a finite  data matrix $\bsbX$  and an estimator  $(\hat \bsbA, \hat\bsbS, \hat\bsbV, \hat\bsbV_{\perp})(\bsbX)$, abbreviated as $\hat \bsbB$, its breakdown point is
$
\epsilon^*(\hat \bsbB)= \frac{1}{n}\cdot\min\{q\in \mathbb Z^+: \sup_{\bsb{X}\in \mathcal {B}( q)}  \EE[ M(\hat \bsbB;\bsbB)] =+\infty\},
$
where $\mathbb Z^+= \mathbb{N}\cup\{0\}$, $\mathcal B(q)=\{\bsbX \in \mathbb R^{n\times p}: \bsbX = \bsbB + \bsbE, \mbox{ where } \bsbB=\bsbA {\bsbV}^{T} + \bsbS \bsbV_{\perp}^{ T}, \vect(\bsbE) \mbox{ is sub-Gaussian},  \bsbA\in \mathbb R^{n\times r}, \|\bsbS\|_{2,0} \leq q,  [\bsbV, \bsbV_{\perp}] \in \mathbb O^{p\times p}\}$.  Note that the randomness of $\hat \bsbB$ is accounted by taking the expectation. %Different with the worst-case finite-sample breakdown point \citep{donoho1983notion}, this gives a \textbf{stochastic} characterization of the breakdown of an estimator.
It follows from   \eqref{oracleineq} that     $\epsilon^*(\hat \bsbB)\geq (q+1)/n$. % (by setting $\bsbS = \bsbS^*$, $\bsbA = \bsbA^*$, $\bsbV=\bsbV^*$).
% the risk of the ROC-PCA estimator is bounded. for any $\|\bsbS\|_{2,0}\le q$ and so

Furthermore, we show that in a minimax sense, the  error rate obtained in  Theorem \ref{mytheorem:oracleerr}  is essentially optimal.
Consider the following  signal class
 \begin{align}
\mathcal {S}(r, q)= \{(\bsbA^*, \bsbS^*, \bsbV^*, \bsbV_{\perp}^*): \bsbA^*\in \mathbb R^{n\times r}, [\bsbV^*, \bsbV_{\perp}^*] \in \mathbb O^{p\times p}, \|\bsbS^*\|_{2,0}\leq q   \},
\end{align}
where ${1} \leq q \leq n$, ${1} \leq r \leq n\wedge p$.
Let $\ell(\cdot)$ be a nondecreasing loss function with $\ell(0)=0$, $\ell \not\equiv0$. %We show that a minimax lower bound holds for any estimator $(\hat\bsbA, \hat\bsbS, \hat\bsbV, \hat\bsbV_{\perp})$.
  %This is usually required in minimax studies \citep{Loun2011,bunea2012joint,raskutti2012minimax}.

\begin{theorem}
\label{th_minimax}
Assume $\bsb{X}=\bsbA^*\bsbV^{* T} + \bsbS^* \bsbV_{\perp}^{* T}+\bsbE$ with $e_{ij}\overset{iid} \sim N({0},\sigma^2)$, $n\geq 2$, $1\leq q \leq n$, $1\leq r \leq n\wedge p$, $r(n+p-r)\geq 8$, $q d\geq 8$.  Then there exist positive constants $C$, $c$ (depending on $\ell(\cdot)$ only) such that
\begin{equation}
    \inf_{(\hat\bsbA, \hat\bsbS, \hat\bsbV, \hat\bsbV_{\perp})}\,\sup_{(\bsbA^*, \bsbS^*, \bsbV^*, \bsbV_{\perp}^*) \in \mathcal {S}(r, q)}\mathbb{E}[\ell(M(\hat \bsbA, \hat\bsbS, \hat\bsbV, \hat\bsbV_{\perp};\bsbA^*, \bsbS^*, \bsbV^*, \bsbV_{\perp}^*)/(C P_o(q,r)))] \geq c>0, \label{minimaxlowerbound}
\end{equation}
where   $(\hat\bsbA, \hat\bsbS, \hat\bsbV, \hat\bsbV_{\perp})$ denotes an arbitrary  estimator of $(\bsbA^*, \bsbS^*, \bsbV^*, \bsbV_{\perp}^*)$ and $P_o(q, r) = P_o(q, r;  n, p, \sigma^2)={ \sigma^2} \{ q d + r n + r p +   q\log(en/q) \}/({n p})$.
\end{theorem}

  We give some examples of $\ell$ to illustrate the conclusion. Using the indicator function $\ell(u)=1_{u\geq 1}$, we learn that for any estimator  $(\hat\bsbA, \hat\bsbS, \hat\bsbV, \hat\bsbV_{\perp})$,
$
M(\hat \bsbA, \hat\bsbS, \hat\bsbV, \hat\bsbV_{\perp};\bsbA^*, \bsbS^*, \bsbV^*, \bsbV_{\perp}^*) \gtrsim \sigma^2\{r(p +n ) + qd + q\log(en/q)\}/(np)
$
occurs with positive probability. For $ \ell(u)=u$, the theorem shows that the risk $\mathbb{E}[M(\hat \bsbA, \hat\bsbS, \hat\bsbV, \hat\bsbV_{\perp};\bsbA^*, \bsbS^*, \bsbV^*, \bsbV_{\perp}^*)]$ is bounded from below by the same rate % $P_o(q, r)$
 up to some multiplicative constant. Therefore, ROC-PCA can essentially achieve the minimax  optimal rate non-asymptotically. % Moreover, our  rate analysis is nonasymptotic and applies to any $n, p, m$.

%\item

Various asymptotic results can be obtained from the finite-sample bound. In fact, as long as $np\gg q p + r n + r p +   q\log(en/q)$, the approximation error tends to zero.  In real-life applications, the values of   $q$ (the number of outliers) and $r$ (the number of principal components) of interest are typically very small (even as small as 2 or 1), in which case  the proposed  ROC-PCA,  exploiting the parsimony offered by  low rankness \textit{and} sparsity, has  guaranteed  small error in theory.

%\item In the case where element-wise sparsity in $\bsbS$ is assumed, i.e.,  $\|\mbox{vec}(\bsbS)\|_{0}\leq q^e$, we can show a similar theorem with $P_0^e$ as the error rate:
%\begin{align}
%P_o^e(r, q^e, n, p, \sigma^2)=\frac{ \sigma^2}{n p} \{ q^e + nr + r p +   q^e\log(e n p/q^e) \}.
%\end{align}
%\end{enumerate}

\section{Batch ROC-PCA}
\label{sec:srocpca_rocpca}
Modern  applications call for the need of scalable algorithms in high dimensions. Unfortunately, most methods reviewed in previous sections suffer from heavy computational burden when directly applied to  $p\gg n$ datasets (and some may fail in principle). A widely acknowledged practice in the   robust PCA literature is to perform an SVD reduction beforehand  (see, e.g., \cite{hubert2002fast}).  %\cite{hubert2005robpca}.
Nevertheless, we found that such a pre-processing may be unreliable and non-robust when $p$ is very high. In this section, we discuss it in details and propose a {batch} variant of ROC-PCA to meet the challenge.
% Our batch ROC-PCA can essentially be applied to data of any sizes.

\paragraph{SVD reduction}
%\label{sec:svdpreprocessing_rocpca}
People usually conduct an SVD reduction in advance before applying robust PCA to high dimensional data: given $\bsbX$ with all its columns properly centered, obtain its top $\mathfrak q$ right singular vectors in $\bsbV^{o}$, with $\mathfrak q=\text{rank}(\bsbX)$ typically, and form $\bsbtX_{n\times \mathfrak q} = \bsbX_{n\times p}\bsbV^o_{p\times \mathfrak q}$. Then, apply  robust PCA  on $\bsbtX$ with the resultant estimate denoted by $\bsbtV \in \mathbb O^{\mathfrak q \times r}$. In the end, $\bsbV^o\bsbtV$ is reported as the estimated PC directions. In such a  procedure,  the computational burden of robust PCA can be significantly reduced.

{The  SVD reduction is commonly believed to be valid for robust PCA and does not cause  any information loss (from which  it appears that the challenge of  high dimensionality is not that serious).} But it just amounts to a rank-$\mathfrak q$ PCA. In fact, even assuming (ideally) that the true column means can be accurately and robustly estimated, the  obtained directions may be misleading when $p\gg n$ and/or in the presence of OC outliers.

If the back-transformed estimate $\bsbV^o\bsbtV$ coincides with the authentic loading matrix $\bsbV^*$, then the  PC subspace
%$\bsbP_{\bsbV^*}:=\bsbV^*\bsbV^{*T}$
 must lie in the observed row space% $\bsbP_{\bsbV^o}:=\bsbV^o\bsbV^{oT}$
, namely, $\bsbP_{\bsbV^*} \subseteq \bsbP_{\bsbV^o}$. Hence the belief in the reduction %in performing the SVD reduction
 is that as long as $\bsbX\bsbu$ is $\bsb{0}$, or approximately so, $\bsbu$ should not contain much information about $\bsbP_{\bsbV^*}$ (or deserve to be checked for OC outliers).

 Let's consider a toy example with the $i$-th row of $\bsbX^*$ given by $[a_i,0,\cdots,0]$, and the $i$-th row of the corrupted matrix $\bsbX$ given by  $[a_i,\epsilon a_i,\cdots,\epsilon a_i]$, where  $\epsilon$  is set small enough. % so that $[1, 0, \cdots, 0]^T$ is the dominant PC direction.
Then, we have  $\bsbX\bsba=\bsb{0}$ for $\bsba =[1,-\frac{1}{\epsilon(p-1)},\cdots,-\frac{1}{\epsilon(p-1)}]^T$. With  $p$  ultra-high, $\bsba$  and   $[1, 0, \cdots, 0]^T$ determine nearly the same projection, indicating that   the true PC subspace essentially  lies in the  orthogonal complement of the observed row space!
%Not only that the observed row space $\bsbP_{\bsbV^o}$ may not fully cover the true PC subspace $\bsbP_{\bsbV^*}$, but $\bsbP_{\bsbV^*}$ may lie in the orthogonal complement of $\bsbP_{\bsbV^o}$.
Accordingly, simply applying the SVD reduction is questionable and the curse of dimensionality is non-trivial. This perhaps surprising finding is  closely connected to \cite{johnstone2009consistency}. It is easy to show that the existence of OC outliers only makes this phenomenon much more severe.

Therefore, we caution against such a plain PCA based dimension reduction in ultra-high dimensional problems (with possible OC outliers). On the other hand, a reduction can be safely made in the OC subspace with the help from ROC-PCA. % from $p$ to $p^{\prime}$ with $r \le p^{\prime}<p$, as will be explained below. This observation is quite useful in the design of batch ROC-PCA.

\paragraph{Batch ROC-PCA}
 We propose a batch ROC-PCA (BROC-PCA) to speed the computation. The basic idea is to estimate $\bsbVp$ in a \emph{batch} fashion. Each  time,  identify only $m$ ($m<d$) least significant OC loading vectors. By setting $m \ll p/2$, the  inversion formula based update in Section~\ref{sec:vpopt_rocpca} can be effectively used. Moreover, a rank-$(p-m)$ SVD reduction can be  performed afterwards as  the reduced $m$ least significant dimensions contain no PC information. This step makes the problem size  drop after each batch processing.
% the deflated working matrix contains no more outliers in the $m$ least significant dimensions

Concretely, given $\bsbX_1:=\bsbX$ and a series of batch sizes $m_k$ ($1 \le k \le K$) satisfying $\sum_{k=1}^K m_k = d$, the  BROC-PCA procedure is as follows. For each $k$, apply ROC-PCA to $\bsbX_k$ with the resultant estimate denoted by $\bsbV_{\perp,k}$,  containing $m_k$ least significant OC loading vectors of $\bsbX_k$. Form an intermediate matrix $\bsbZ=\bsbX_k(\bsbI-\bsbV_{\perp,k}\bsbV_{\perp,k}^T)$, and obtain $\bsbX_{k+1} = \bsbZ\bsbV_k$, where $\bsbV_k$ is the top $(p-\sum_{i=1}^k m_i)$ right singular vectors of $\bsbZ$. Finally, the product of all $\bsbV_1,\cdots,\bsbV_K$ is delivered as the PC directions estimate, i.e., $\hat{\bsbV}: = \prod_{k=1}^K \bsbV_k$.
%\underline{\textbf{I did not write $\bsbS$ update for now.}}
% , and $\bsbS_k$ is the $n\times m_k$ outlier matrix
% In addition, to get the final outlier matrix estimate $\hat{\bsbS}$, first concatenate all $\bsbS_k$ to form $[\bsbS_1,\cdots,\bsbS_K]$, and then keep the rows with the largest $q$ row norms, and set all other rows as $\bsb{0}$.
An attractive feature  is that the number of columns of $\bsbX_k$ gets smaller as $k$ increases. In implementation, to take advantage of the fast update formula, we recommend choosing $m_k$ satisfying $m_k < (p-\sum_{i=0}^{k-1} m_i)/2$ (assuming $m_0=0$) unless the problem size is sufficiently small. A rule of thumb in large-$p$ computation is $30 \le m_k \le 100$. To attain further speedup, we adopt a progressive error control scheme---the error tolerance used in the ROC-PCA algorithm is gradually tightened up from the computation of the first batch to the $K$-th batch. % after calculating the $k$-th batch of OC loadings.

BROC-PCA shares similarity with some  sparse PCA algorithms, e.g., \cite{shen2008sparse}. % and \cite{witten2009penalized}.
However, instead of repeatedly solving the rank-1  problem, BROC-PCA estimates $m_k$ OC loading vectors at each time;  more importantly,  the SVD reduction is then  employed to reduce the dimensionality  by $m_k$. Accordingly, the overall computational cost can be substantially reduced (by  about 70\% for $p=1000$).

\section{Simulations}
\label{sec:simulation_rocpca}

\subsection{Choice of \ensuremath{q}}
\label{sec:simmulationQtuning_rocpca}
We first  study how the parameter $q$, as an upper bound of the number of outliers, affects the performance of ROC-PCA in subspace recovery. We generate observations  according to the model  $\bsbX = \bsbU \bsbD \bsbV^T + (\bsb{1}\bsbmu^T+\bsbS)\bsbV_{\perp}^{T}+\bsbE$ with random orthogonal $\bsbU$ and $\bsbV$. In this experiment,  $n=100$, $p=10$, $r=3$, $\bsbD=\mbox{diag}\{60,40,20\}$, $\bsbmu = \bsb{0}$ and $\sigma^2=2$. The outlier matrix $\bsbS$ (assuming r-Type outliers for now) has the first $O$ rows as $L\cdot [1,\cdots,1]$ and $\bsb{0}$ otherwise. We consider different combinations of the number of outliers $(O)$ and leverage value $(L)$ with $O \in \{4,10,16\}$ and $L \in \{3.5,4.5\}$. In calling the ROC-PCA algorithm, we try $q=\alpha O$ with $\alpha$ varying over the grid $\{0.8,1,1.5, 2, 2.5,3,3.5,4\}$. Each model is simulated 50 times, and the mean results of subspace estimation and outlier identification are reported using the following metrics.

1) \emph{Subspace estimation}. The accuracy of any PC subspace estimate is measured by the cosine value of the largest \emph{canonical angle} (denoted by $\theta$) between $\bsbP_{\hat{\bsbV}}:= \hat{\bsbV}\hat{\bsbV}^T$ and $\bsbP_{\bsbV^*}:=\bsbV^*\bsbV^{*T}$, where $\hat{\bsbV}$ is an estimate of the ideal top $r$ PC loadings $\bsbV^*$. It is well known that  $\|\bsbP_{\hat{\bsbV}}-\bsbP_{\bsbV^*}\|_2 = \sqrt{1-\cos ^2 \theta},$ where $\|\cdot\|_2$ denotes the spectral norm, see, e.g.,~\cite{gohberg1969introduction}. In the rest of the paper, $100\cdot\cos \theta$ and \emph{PC affinity} are used interchangeably to evaluate the subspace estimation accuracy.
% Ideally, $\cos \theta \approx 1$.\

2) \emph{Outlier identification}. Three benchmark measures are used: the mean masking (\textbf{M}) and swamping (\textbf{S}) probabilities, and the rate of successful joint detection (\textbf{JD}). The masking probability is the fraction of undetected outliers, the swamping probability is the fraction of good points labeled as outliers, and the JD is the fraction of simulations with zero masking. An ideal method should have $\text{M} \approx 0\%$, $\text{S} \approx 0\%$, and $\text{JD} \approx 100\%$. However, for estimation purposes, masking is a much more serious problem than swamping, because the former can cause serious distortion, whereas the latter is often just a matter of losing efficiency.

\begin{figure}[!h]
    \centering
    \subfigure{
        \begin{minipage}[b]{0.45\textwidth}
            \includegraphics[scale=0.45,clip]{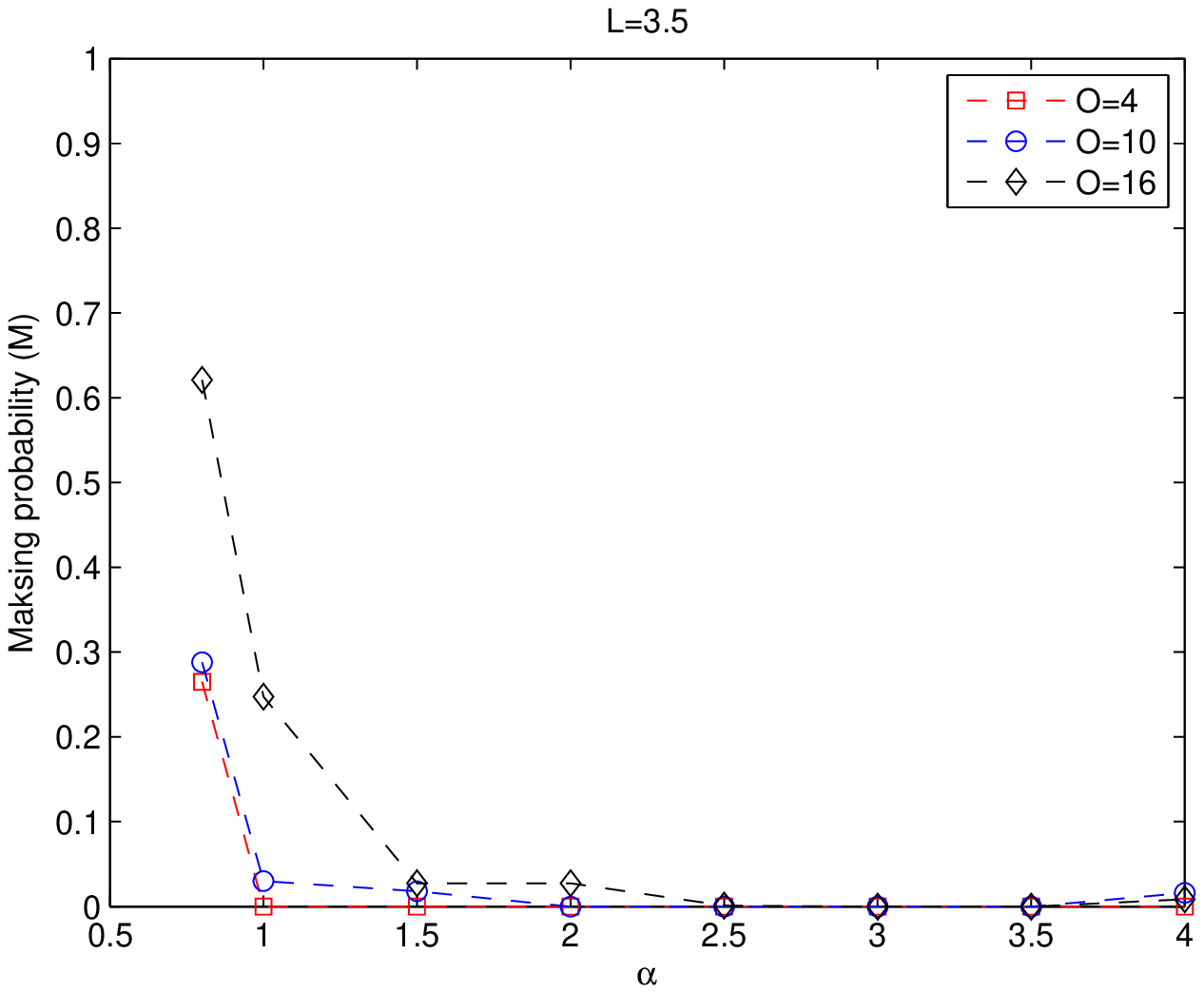}
            \includegraphics[scale=0.45,clip]{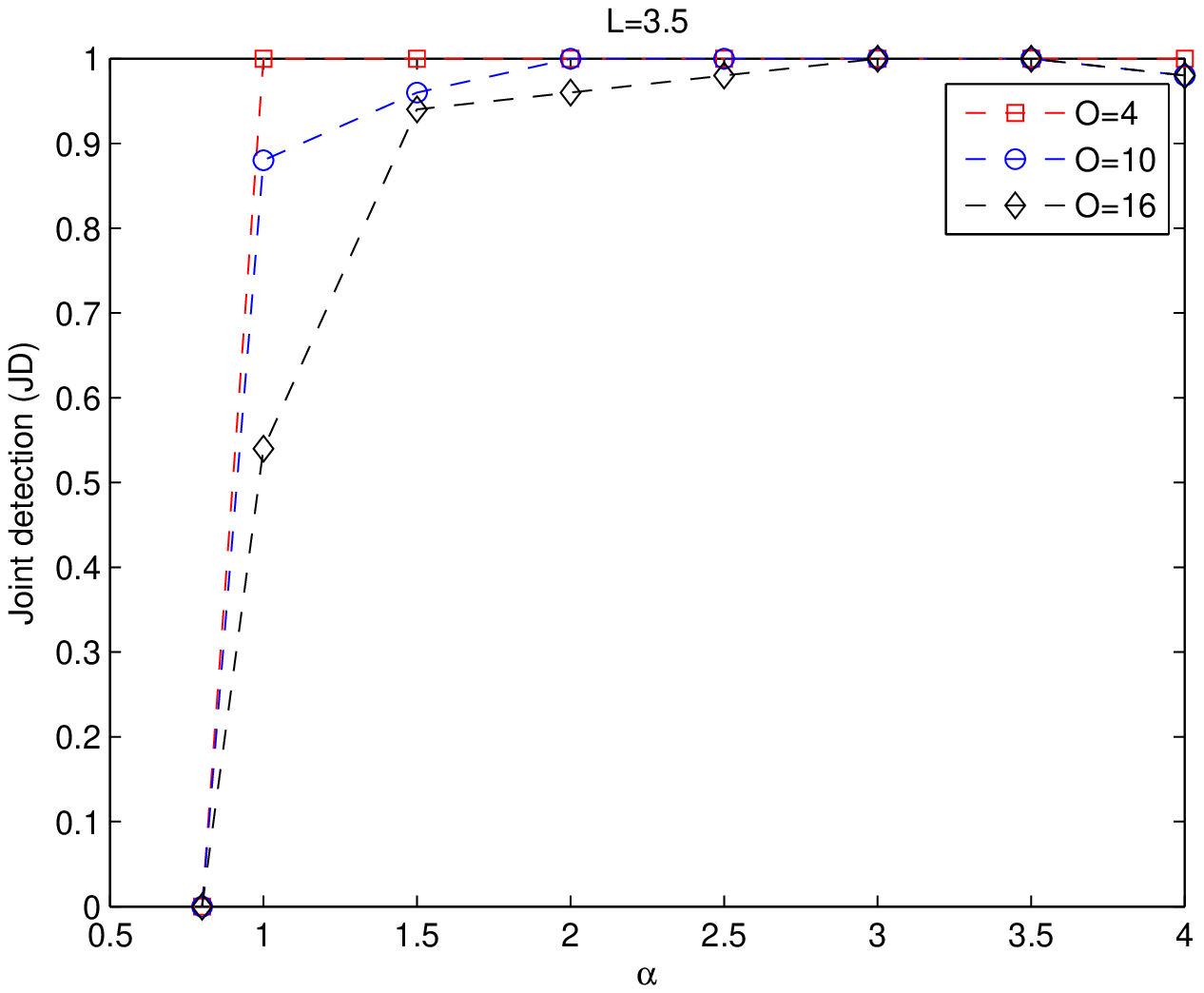}
            % \caption{$L=3.5$}
        \end{minipage}

        \hspace{-0.2in}

        \begin{minipage}[b]{0.45\textwidth}
            \includegraphics[scale=0.45,clip]{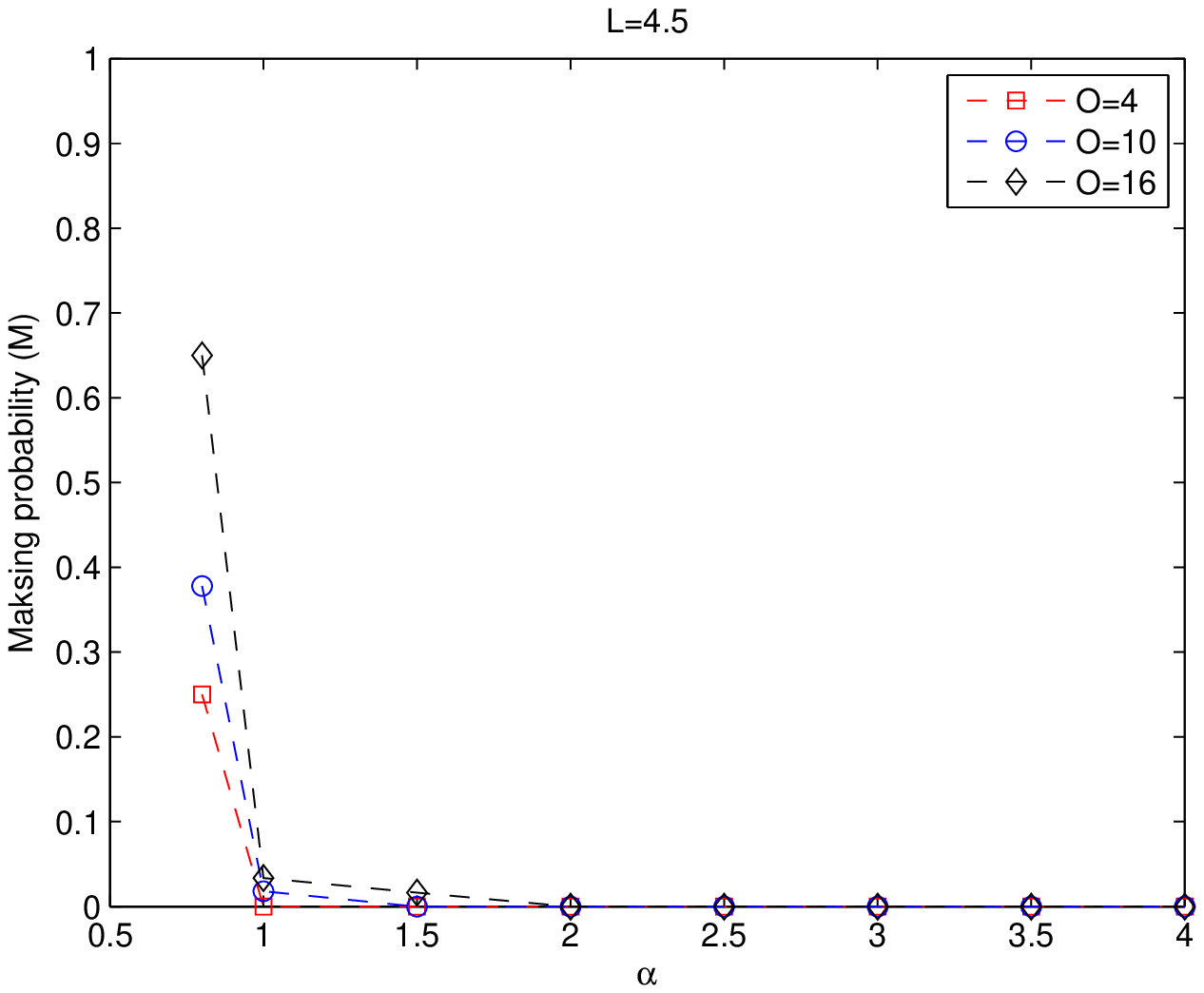}
            \includegraphics[scale=0.45,clip]{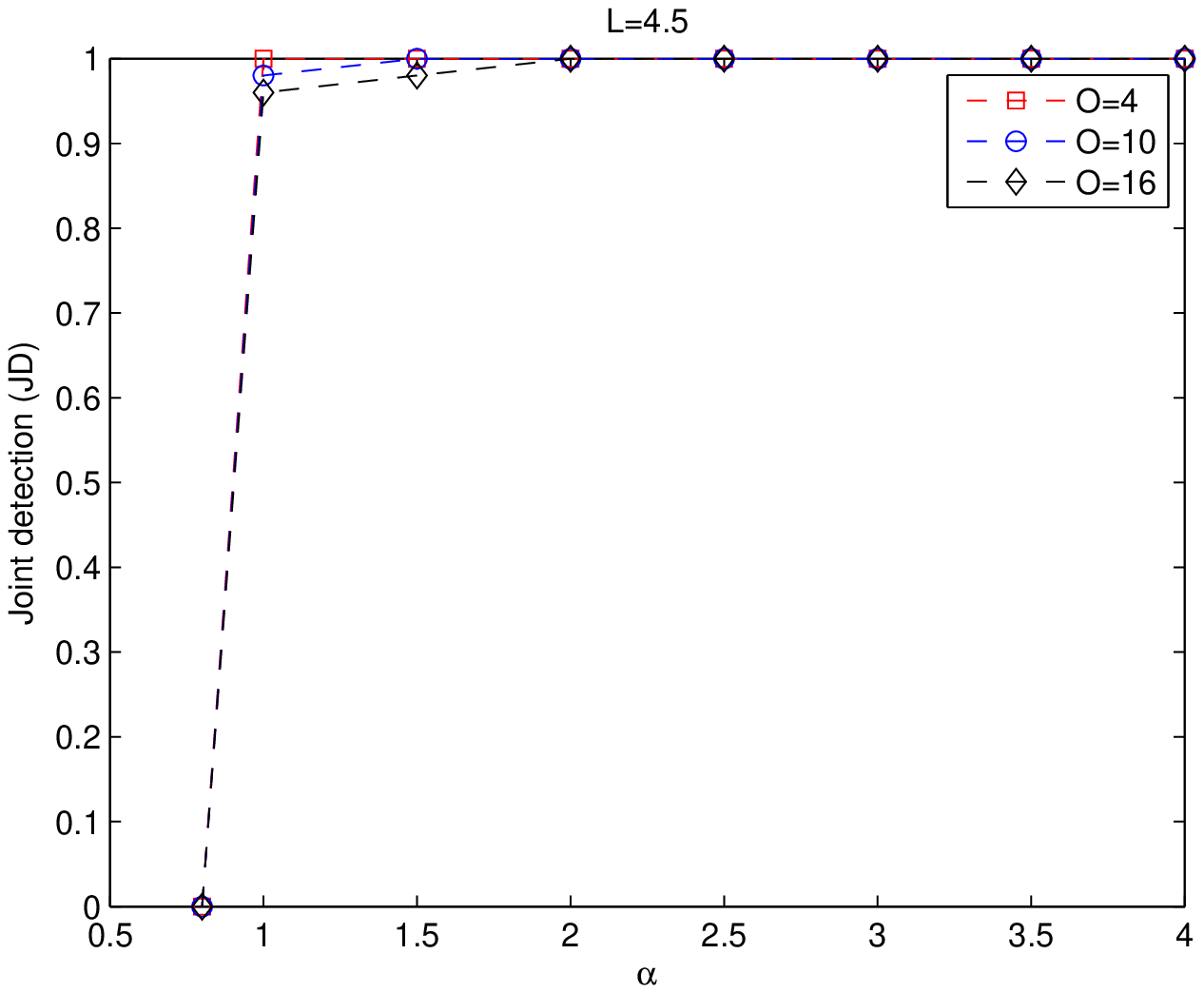}
            % \caption{$L=3.5$}
        \end{minipage}
    }
    % \subfigure{\includegraphics[scale=0.5,clip]{./figure/qtuningML3_5.eps}}
    % \subfigure{\includegraphics[scale=0.5,clip]{./figure/qtuningJDL3_5.eps}}
    \caption{\small Masking (M) and joint detection (JD) results with respect to $\alpha$ on simulation data with different $(O$, $L)$ combinations.}%  of the number of outliers $(O)$ and leverage value $(L)$ $\alpha$ directly affects the upper outlier number bound parameter $q$ by $q=\alpha O$.
    \label{fig:qtuning}
\end{figure}

The simulation results are summarized in Table~\ref{tab:sim_qtuning_performance}. To provide some intuition for choosing the parameter, we also plot  in Figure~\ref{fig:qtuning} the rates of M and JD  with respect to $\alpha$ under different $(O$, $L)$ combinations. When the value of $q$ is smaller than the true number of outliers, masking always occurs and may seriously affect the performance of subspace estimation (e.g., $\alpha=0.8$ and $O=16$). But as  $q$ increases, the masking probability decreases dramatically. Indeed, as long as the value of $q$ exceeds the true number of outliers, i.e., $\alpha > 1$, ROC-PCA nearly achieves  ideal M and JD results, even when the outliers have high leverage values (such as  $L=4.5$). An important finding is that  the subspace estimation accuracy does not change much when $\alpha$ is in a relatively large   range, say, $1<\alpha \le 3$. Such an insensitivity greatly facilitates $q$-tuning. Of course, if $q$ is set too big, the swamping effect becomes  serious and can affect estimation efficiency. (For instance, when $\alpha=4$, $q=64$ out of $100$ observations have to be removed in estimation.) In general, ROC-PCA performs better in masking, but slightly worse in swamping. This is an acceptable trade-off, because masking is more harmful in robust estimation.   $\alpha=2$ seems to be a good choice to strike a balance.
% This indicates that the estimation performance is insensitive to $q$ when $1<\alpha \le 3$.

\begin{table}[H]% of the number of outliers $(O)$ and leverage value $(L)$
\setlength{\tabcolsep}{4pt}\centering \caption{\small Subspace estimation and outlier identification  of ROC-PCA for  different choices of $\alpha$ (or $q$) on simulation data with various $(O$, $L)$ combinations.} \label{tab:sim_qtuning_performance}
\footnotesize%\scriptsize%\tiny
  \begin{tabular}{c  c  c c c c  c c c c  c c c c}
  \multicolumn{11}{c}{}\\
    \hline \hline
    \multirow{2}{*}{$L$} & \multirow{2}{*}{$\alpha$}& \multicolumn{4}{c}{$O=4$}  & \multicolumn{4}{c}{$O=10$} &  \multicolumn{4}{c}{$O=16$} \\ \cmidrule(r){3-6} \cmidrule(r){7-10} \cmidrule(r){11-14}
    & & Affinity & M & S & JD & Affinity & M & S & JD  & Affinity & M & S & JD  \\ \hline

    \multirow{7}{*}{4.5}
& $0.8$ & 96 & 0.250 & 0.000 & 0.000   &   72 & 0.378 & 0.020 & 0.000   &   24 & 0.650 & 0.076 & 0.000 \\
& $1.0$ & 97 & 0.000 & 0.000 & 1.000   &   95 & 0.018 & 0.002 & 0.980   &   93 & 0.034 & 0.006 & 0.960 \\
& $1.5$ & 97 & 0.000 & 0.021 & 1.000   &   97 & 0.000 & 0.056 & 1.000   &   95 & 0.016 & 0.098 & 0.980 \\
& $2.0$ & 97 & 0.000 & 0.042 & 1.000   &   96 & 0.000 & 0.111 & 1.000   &   95 & 0.000 & 0.190 & 1.000 \\
& $2.5$ & 96 & 0.000 & 0.063 & 1.000   &   96 & 0.000 & 0.167 & 1.000   &   93 & 0.000 & 0.286 & 1.000 \\
& $3.0$ & 97 & 0.000 & 0.083 & 1.000   &   96 & 0.000 & 0.222 & 1.000   &   92 & 0.000 & 0.381 & 1.000 \\
& $3.5$ & 96 & 0.000 & 0.104 & 1.000   &   95 & 0.000 & 0.278 & 1.000   &   89 & 0.000 & 0.476 & 1.000 \\
& $4.0$ & 96 & 0.000 & 0.125 & 1.000   &   94 & 0.000 & 0.333 & 1.000   &   82 & 0.000 & 0.571 & 1.000 \\  \hline

    \multirow{7}{*}{3.5}
& $0.8$ & 95 & 0.265 & 0.001 & 0.000   &   85 & 0.288 & 0.010 & 0.000   &   37 & 0.621 & 0.071 & 0.000 \\
& $1.0$ & 97 & 0.000 & 0.000 & 1.000   &   95 & 0.030 & 0.003 & 0.880   &   72 & 0.248 & 0.047 & 0.540 \\
& $1.5$ & 97 & 0.000 & 0.021 & 1.000   &   95 & 0.018 & 0.058 & 0.960   &   93 & 0.028 & 0.100 & 0.940 \\
& $2.0$ & 97 & 0.000 & 0.042 & 1.000   &   96 & 0.000 & 0.111 & 1.000   &   92 & 0.028 & 0.196 & 0.960 \\
& $2.5$ & 97 & 0.000 & 0.063 & 1.000   &   96 & 0.000 & 0.167 & 1.000   &   93 & 0.001 & 0.286 & 0.980 \\
& $3.0$ & 96 & 0.000 & 0.083 & 1.000   &   95 & 0.000 & 0.222 & 1.000   &   93 & 0.000 & 0.381 & 1.000 \\
& $3.5$ & 96 & 0.000 & 0.104 & 1.000   &   94 & 0.000 & 0.278 & 1.000   &   88 & 0.000 & 0.476 & 1.000 \\
& $4.0$ & 96 & 0.000 & 0.125 & 1.000   &   91 & 0.016 & 0.335 & 0.980   &   86 & 0.009 & 0.573 & 0.980 \\ \hline

  \end{tabular}
\end{table}

\subsection{Comparisons with some existing methods}
\label{sec:simmulationComparison_rocpca} We focus on the subspace estimation accuracy and compare  ROC-PCA with the plain PCA and some representative robust PCA methods reviewed earlier, including RAPCA, ROBPCA, spherical PCA (S-PCA) and PCP. Note that the  fast-MCD, as one of the most popular  robust covariance matrix based methods, is already integrated in ROBPCA.  RAPCA and ROBPCA  are provided by the library {LIBRA} \citep{verboven2005libra}.
 %In running ROBPCA, all parameters are set to be their default values, unless otherwise specified.
  Our implementation of  S-PCA  follows \cite{locantore1999robust}. For PCP, we use the inexact augmented Lagrange multiplier algorithm \citep{lin2010augmented}. (We also tested  the stable PCP  and the  EGMS algorithms  \citep{zhou2010stable,zhang2014novel}, but  they do not show better statistical performances  than  PCP in the presence of OC outliers, and  take more time. So   PCP is  used as a representative  method in the class of low-rank  approximation based robust PCA.)
% with the default parameter settings.

\paragraph{Comparison in the presence of  r-Type outliers}
We consider two setups with $n>p$ and $n<p$, respectively. In both setups, the data matrix $\bsbX$ is generated according to  Section~\ref{sec:simmulationQtuning_rocpca}, with $r=3$, $\bsbD=\mbox{diag}\{100,60,20\}$, $\bsbmu=\bsb{0}$ and $\sigma^2 \in \{0.5,1\}$. The first $O$ rows of the outlier matrix $\bsbS$ are all given by $[10,\cdots,10]$, while the rest rows are  $\bsb{0}$. In the $n>p$ situation, $n=100$, $p =50$, and $O \in \{4,10,16\}$. In the case of $n<p$, $n=50$, $p=100$, and $O \in \{2,5,8\}$. The major concern here is the subspace estimation accuracy and we use a conservative choice  $q=2O$ for breakdown. % in ROC-PCA and ROBPCA.  (In this way, ROBPCA behaves mildly better than using the default value).

\begin{table}[H]
\centering
\caption{\small PC affinity comparison in the presence of r-Type OC outliers.}
 \label{tab:sim_rType} \footnotesize% \scriptsize
  \begin{tabular}{c | c | c | c c c c c c }
  \multicolumn{9}{c}{} \\
          \hline
          \hline            %$(n,p,r)=(100,5,3)$
        $(n,p)$ & $\sigma^2$ & $O$ & PCA    & PCP & S-PCA    & RAPCA  & ROBPCA  &  \textbf{ROC-PCA}\\ \hline

\multirow{6}{*}{$(100,50)$}
& \multirow{3}{*}{0.5} & $4$   &   0 &   6 &  69 &  51 &  96 &  96\\
&                      & $10$ &   0 &   1 &   3 &  53 &  96  &  96 \\
&                      & $16$  &   0 &   1 &   1 &  33 &  95 &  95\\

\cline{2-9}
& \multirow{3}{*}{1}   & $4$   &   3 &   7 &  19 &  31 &  92 &  92 \\
&                      & $10$ &   1 &   1 &   3 &  27 &  92  &  92\\
&                      & $16$  &   0 &   2 &   1 &  15 &  92 &  90 \\
\hline

\multirow{6}{*}{$(50,100)$}
& \multirow{3}{*}{0.5} & $2$   &   1 &   3 &  90 &  31 &  94 &  94\\
&                      & $5$   &   0 &   2 &   3 &  21 &  93 &  93\\
&                      & $8$   &   2 &   1 &   3 &  35 &  92 &  92 \\
\cline{2-9}

& \multirow{3}{*}{1}   & $2$   &   1 &   2 &  60 &  21 &  87 &  87 \\
&                      & $5$   &   0 &   1 &   2 &  20 &  85 &  85 \\
&                      & $8$   &  1 & 1 & 1 & 18  & 84 &  84\\
\hline

$(450,15)$ & 0.001     & $2$   &  0 & 100 & 100 & 100 & 1 & 100\\ \hline
  \end{tabular}
\end{table}

\begin{table}[H]
\centering
\caption{\small Computational time (in seconds) comparison in the presence of r-Type OC outliers.}
 \label{tab:sim_time_rType} \footnotesize% \scriptsize
  \begin{tabular}{c | c | c | c c c c c c }
  \multicolumn{9}{c}{} \\
          \hline
          \hline            %$(n,p,r)=(100,5,3)$
        $(n,p)$ & $\sigma^2$ & $O$ & PCA    & PCP & S-PCA    & RAPCA  & ROBPCA &  ROC-PCA \\ \hline

\multirow{6}{*}{$(100,50)$}
& \multirow{3}{*}{0.5} & $4$   &   0.02 &   0.12 &  0.03 &  0.14 &  1.06 &  5.00\\
&                      & $10$  &   0.01 &   0.11 &   0.02 &  0.12 &  1.03 &  10.3\\
&                      & $16$  &   0.01 &   0.11 &   0.02 &  0.12 &  1.08 &  15.9\\

\cline{2-9}
& \multirow{3}{*}{1}   & $4$  &   0.01 &   0.11 &  0.02 &  0.12 &  0.98  &  4.27 \\
&                      & $10$ &   0.01 &   0.11 &   0.02 &  0.12 &  1.01 &  18.2 \\
&                      & $16$  &   0.01 &   0.11 &   0.02 &  0.13 &  1.05 &  35.9 \\
\hline

\multirow{6}{*}{$(50,100)$}
& \multirow{3}{*}{0.5} & $2$   &   0.03 &   0.16 &  0.03 &  0.07 &  0.94&  13.4 \\
&                      & $5$   &   0.01 &   0.14 &   0.02 &  0.06 &  0.83 &  20.5\\
&                      & $8$   &   0.01 &   0.16 &   0.03 &  0.05 &  0.83 &  37.0\\
\cline{2-9}

& \multirow{3}{*}{1}   & $2$   &   0.01 &   0.16 &  0.03 &  0.05 &  0.83 &  10.8\\
&                      & $5$   &   0.01 &   0.14 &   0.02 &  0.05 &  0.88 &  23.2 \\
&                      & $8$  &  0.01 & 0.15 & 0.02 & 0.05  & 0.81  &  29.9\\
\hline

$(450,15)$ & 0.001     & $2$   &  0.05 & 0.05 & 0.04 & 1.19 & 1.15 & 2.63\\ \hline
  \end{tabular}
\end{table}

According to the results of PCA in Table \ref{tab:sim_rType}, the existence of OC outliers can severely skew the estimated PC directions.   PCP did not show much improvement in handling OC outliers.
%, because it targets at elementwise  outliers in the observation space and may not  be able to handle serious OC outliers.
 RAPCA behaved  better than PCA and PCP. But its performance deteriorates when  the noise level increases. The performance of S-PCA is not stable; in particular,  it showed poor results   with $\geq 10\%$ outliers occurring. In contrast,  ROC-PCA and ROBPCA   behaved excellently  in the aforementioned two setups.

To make the story complete, consider   $n=450$, $p=15$, $O=2$, $\sigma^2 = 0.001$ as shown at the bottom  of Table~\ref{tab:sim_rType}.     In this extremely simple case, all robust PCA approaches perfectly estimated the PC subspace except  ROBPCA which gave an  extremely low value of PC affinity. Unfortunately, such a limitation of ROBPCA  is  commonly observed for datasets with large $n$, small $O$ and low $\sigma^2$.  It is largely due to the failure of  the trial direction sampling. (Indeed, in this situation most trial directions used in the multi-step ROBPCA fall into the PC subspace and so  there is a large chance for outliers  to remain in the  $h$-subset---see Section \ref{sec:survey_rocpca}.)

The computational times are reported in  Table \ref{tab:sim_time_rType}. ROC-PCA did not run  as fast as the other procedures/algorithms,  but was definitely affordable.  This  is an acceptable tradeoff between performance and computational complexity in  robust PCA. Of course, faster algorithms are still in great need, which is  an interesting topic for further research.

\paragraph{Comparison in the presence of e-Type outliers}
To test for e-Type outliers, we generated the data under $n=100$, $p=18$, $r=3$, $\bsbD=\mbox{diag}\{80,60,40\}$, $\bsbmu=\bsb{0}$, $\sigma^2 \in \{0.5,1\}$. The outlier matrix $\bsbS$ has $O^e  \in \{60,120\}$ random entries set to be $15$ and $0$ otherwise. Again, we set $q^e=2O^e$ following the discussion in Section~\ref{sec:simmulationQtuning_rocpca}. Table~\ref{tab:sim_eType} and Table \ref{tab:sim_time_eType} show the mean  PC affinity values and mean computational times  from  $50$ runs. Although ROC-PCA runs slower than the other methods, it  shows superior statistical performance, especially when the number of outliers is large.

\begin{table}[H]
\centering
\caption{\small PC affinity comparison for \textbf{e-Type} OC outliers.}
 \label{tab:sim_eType} \footnotesize%\scriptsize
  \begin{tabular}{c | c | c c c c c c }
  \multicolumn{6}{c}{} \\
          \hline
          \hline
        $\sigma^2$ & $O^e$  & PCA    & PCP & S-PCA    & RAPCA  & ROBPCA &  \textbf{ROC-PCA}\\ \hline

%  \multirow{2}{*}{0.5} & $60$  & 99.57 & 16.43 & 96.21 & 97.76 & 84.67 & 79.84 \\
%                       & $120$ & 98.79 & 9.24 & 17.33 & 53.96 & 66.74 & 10.30 \\
% \hline
%  \multirow{2}{*}{1}   & $60$  & 99.07 & 19.59 & 93.81 & 96.66 & 86.28 & 78.11 \\
%                       & $120$ & 98.51 & 9.03 & 13.61 & 47.56 & 60.01 & 9.69 \\
 \multirow{2}{*}{0.5} & $60$  & 16 & 97 & 98 & 85 & 80 & 100 \\
                      & $120$  & 9 & 53 & 54 & 67 & 10 & 99\\
\hline
 \multirow{2}{*}{1}   & $60$   & 20 & 94 & 97 & 86 & 78 & 99\\
                      & $120$  & 9 & 47 & 48 & 60 & 10 & 99\\
\hline
  \end{tabular}
\end{table}

\begin{table}[H]
\centering
\caption{\small Computational time  comparison for \textbf{e-Type} OC outliers.}
 \label{tab:sim_time_eType} \footnotesize%\scriptsize
  \begin{tabular}{c | c | c c c c c c }
  \multicolumn{6}{c}{} \\
          \hline
          \hline
        $\sigma^2$ & $O^e$  & PCA    & PCP & S-PCA    & RAPCA  & ROBPCA &  ROC-PCA\\ \hline

%  \multirow{2}{*}{0.5} & $60$  & 99.57 & 16.43 & 96.21 & 97.76 & 84.67 & 79.84 \\
%                       & $120$ & 98.79 & 9.24 & 17.33 & 53.96 & 66.74 & 10.30 \\
% \hline
%  \multirow{2}{*}{1}   & $60$  & 99.07 & 19.59 & 93.81 & 96.66 & 86.28 & 78.11 \\
%                       & $120$ & 98.51 & 9.03 & 13.61 & 47.56 & 60.01 & 9.69 \\
 \multirow{2}{*}{0.5} & $60$   & 0.03 & 0.03 & 0.03 & 0.16 & 0.99 & 29.5\\
                      & $120$  & 0.04 & 0.04 & 0.03 & 0.12 & 0.89 & 39.5\\
\hline
 \multirow{2}{*}{1}   & $60$   & 0.01 & 0.03 & 0.01 & 0.12 & 0.89 & 23.0\\
                      & $120$  & 0.02 & 0.04 & 0.03 & 0.12 & 0.90 & 25.4\\
\hline
  \end{tabular}
\end{table}

\paragraph{Comparison in the presence of observation outliers}
As before, we consider two types of outliers. The  r-Type experiments assume the model $\bsbX = \bsbL + \bsbS+\bsbE$, where $\bsbL$ has rank $r$, given by $\bsbL = \bsbU \bsbD \bsbV^T$ with randomly generated orthogonal $\bsbU$ and $\bsbV$, $\bsbS$ is a sparse component, and $\bsbE$ has i.i.d. $\mathcal{N}(0,\sigma^2)$ entries. We set  $n=100$, $p =50$, $r=3$, $\bsbD=\mbox{diag}\{100,60,20\}$ and $\sigma^2 = 1$. The first $O \in \{4,10,16\}$ rows of $\bsbS$ are all given by $[10,\cdots,10]$, while the rest rows are  $\bsb{0}$. As shown by  Table  \ref{tab:sim_rType_pcp_setup},    we found that ROC-PCA and ROBPCA usually have excellent performances. However,  ROB-PCA is subject to the same trail direction sampling issue as in the r-Type OC outlier case; it can  fail in very easy settings (examples not shown).

\begin{table}[H]
\centering
\caption{PC affinity comparison in the presence of r-Type outliers in the observation space. }
 \label{tab:sim_rType_pcp_setup} \footnotesize
  \begin{tabular}{ c |  c   c   c  c    c  c  }
  \multicolumn{7}{c}{} \\
          \hline
          \hline            %$(n,p,r)=(100,5,3)$
        \multirow{1}{*}{$O$}  & \multicolumn{1}{c}{PCA} & \multicolumn{1}{c}{PCP} & \multicolumn{1}{c}{S-PCA} & \multicolumn{1}{c}{RAPCA}  &
        \multicolumn{1}{c}{ROBPCA} &  \multicolumn{1}{c}{\textbf{ROC-PCA}}\\
       % & & & Aff & Time & Aff & Time & Aff & Time & Aff & Time & Aff & Time & Aff & Time & Aff & Time & Aff & Time \\
\hline
 $4$  &  14  &  13 &   41 & 20 &92 &  92  \\
 $10$  & 10 &  10  &    15  &  29 &  91 &  91  \\
 $16$  &  11  &  11 &    13 &  18  &  90 &  89
 \\ \hline
\end{tabular}
\end{table}

The elementwise case is more complicated.  Table \ref{tab:sim_eType_pcp_setup123} gives some examples. The experiments with e-Type outliers in the observation space use the same model, with $\bsbU$   randomly generated,    $n=100$, $p=18$, $r=3$, $\bsbD=\mbox{diag}\{80,60,40\}$, $\sigma^2 = 1$. The following three setups are considered.    1)   $\bsbV$ is randomly generated and $\bsbS$ has $144$ entries set to be $15$ and $0$ otherwise;
2) $\bsbV=\bsbI[:, 1:3]$ and  the first three columns of the outlier matrix $\bsbS$ have $12$ entries set to be $5$;    3)    $\bsbV$ is randomly generated and $\bsbS$ has $72$ entries set to be $20$ and $0$ otherwise.

\begin{table}[H]
\centering
\caption{PC affinity comparison in the presence of e-Type  outliers in the observation space. }
 \label{tab:sim_eType_pcp_setup123}  \footnotesize %\scriptsize%\tiny %
  \begin{tabular}{ c  | c    c  c   c   c  c  }
  \multicolumn{7}{c}{} \\
          \hline
          \hline            %$(n,p,r)=(100,5,3)$
  & \multicolumn{1}{c}{PCA} & \multicolumn{1}{c}{PCP} &  \multicolumn{1}{c}{S-PCA} & \multicolumn{1}{c}{RAPCA}  &
        \multicolumn{1}{c}{ROBPCA}  &  \multicolumn{1}{c}{\textbf{ROC-PCA}}\\
        \hline
       % & & Aff & Time & Aff & Time & Aff & Time & Aff & Time & Aff & Time & Aff & Time & Aff & Time & Aff & Time \\ \hline
Setting 1 &   75 &  98 &    94 &   90 &  64 &  95 \\
\hline
Setting 2&   99 &   28  &   99 &  95 &   99  &  99  \\ \hline
Setting 3 &   79 &   99 &    97 &  88  &  74 &  79  \\
%  &  79 &  99 &     97 & 88 &   79  &  98 \\
\hline
\end{tabular}
\end{table}

%As reported by Table  \ref{tab:sim_eType_pcp_setup123},
Most methods did not perform uniformly well. S-PCA, though simple, did a very good job.   The affinity value of ROC-PCA is not very high in the last setting. A careful examination of the results shows that  although checking the outlyingness in the OC subspace is  reasonable, the e-Type observation outliers may lead to error propagation in the OC coordinates and so  desire a much larger value of  $q^e$. To fix this, we tried  the $\ell_0$ {penalized} ROC-PCA which uses $P(\bsbS;\lambda)= \frac{\lambda^2}{2} \sum 1_{s_{ij} \ne 0}$ with $\lambda$ as the  threshold parameter. It  falls into the $\Theta$-$P$ framework in Section \ref{sec:musopt_rocpca} and corresponds to the hard-thresholding in implementation \citep{she2012iterative}.  With $\lambda$ taken the universal threshold level $\sigma \sqrt {2 \log (nd)}$ (or   optimally tuned using  the true PC subspace), the PC affinity value of ROC-PCA can be increased to $98$.

\paragraph{BROC-PCA in large-$p$ computation}% $\bsbU$ and $\bsbV$ are random orthogonal matrices.
We compare  the standard ROC-PCA and BROC-PCA in large-$p$ setups. Data samples  are generated in the same way as in Section~\ref{sec:simmulationQtuning_rocpca}, with $n=40$, $r=3$, $\bsbD=\mbox{diag}\{80,60,40\}$, $\bsbmu=\bsb{0}$ and $\sigma^2=1.5$. The first four rows of the outlier matrix $\bsbS$ are all given by $[5,\cdots,5]$, and the remaining ones are $\bsb{0}$. We vary  $p$ in $\{100,300,500,1000\}$. The batch sizes we tried are described as follows:  $m_i=35$, $i=1,2$,  and $m_3=27$ for $p=100$;  $m_1=100$, $m_i=70$,  $i=2,3$, and  $m_4=57$ for  $p=300$; $m_i=100$, $1\leq i \leq 3$,  $m_i=70$, $ i =4, 5$, $m_6=57$ for $p=500$; $m_i=100$, $1\leq i \leq 8$, $m_i=70$, $i=9,10$, and $m_{11}=57$ for $p=1,000$. Table~\ref{tab:sim_sbROCPCA} reports the mean CPU times (in seconds) and estimation accuracy from $20$ runs.

%This  corresponds to $10\%$ r-Type OC outliers among the $40$ observations. We study different problem sizes
% Note that one needs to specify a sequence of batch size numbers in order to apply BROC-PCA. Let  $\{\cdot\}^m$ denote $m$ consecutive values. We use $(\{35\}^2, 27)$ for $p=100$, which means $35$, $35$ and $27$ OC loading vectors are to be estimated sequentially in three batches. The specifications for other problem sizes are: $(100,\{70\}^2,57)$ for $p=300$, $(\{100\}^3,\{70\}^2,57)$ for $p=500$ and $(\{100\}^8,\{70\}^2,57)$ for $p=1000$.

\begin{table}[H]
\centering \caption{\small Computational time and estimation accuracy for  the standard ROC-PCA and BROC-PCA with different dimensions.}
 \label{tab:sim_sbROCPCA} \footnotesize%\scriptsize%
  \begin{tabular}{l  c  r  c  r}
  \multicolumn{5}{c}{} \\
  \hline \hline
  \multirow{2}{*}{} & \multicolumn{2}{c}{ROC-PCA} & \multicolumn{2}{c}{ BROC-PCA}  \\ \cmidrule(r){2-3} \cmidrule(r){4-5}
                                & $100\cdot \cos \theta$ &  time   & $100 \cdot \cos \theta$ &  time \\ \hline
            % $p=100$     &   98.40              & 4.45      & 98.04                & 3.88 \\
            % % $n=40$, $p=200$     &   97.29              & 22.75     & 94.64                & 12.75 \\
            % $p=300$     &   95.37              & 77.07     & 92.73                & 32.78  \\
            % $p=500$     &   91.87              & 265.24    & 88.69                & 95.97  \\
            % $p=1000$    &   88.26              & 2624.40   & 84.24                & 816.83 \\ \hline

            $p=100$     &   98              & 4.5      & 98               & 3.9 \\
            % $n=40$, $p=200$     &   97.29              & 22.75     & 94.64                & 12.75 \\
            $p=300$     &   95              & 77.1     & 93                & 32.8  \\
            $p=500$     &   92             & 265.2    & 89               & 95.9  \\
            $p=1000$    &   88              & 2624.4   & 84               & 816.8 \\ \hline
  \end{tabular}
\end{table}
% As shown in Table~\ref{tab:sim_sbROCPCA}, while BROC-PCA offers comparable estimation accuracy with the standard ROC-PCA, it runs consistently faster, and the reduction of the computational cost is impressive. Even when $p=100$, the speed improvement is available. As $p$ increases, the improvement becomes more evident, for example, when $p=1000$, the time consumption of the standard ROC-PCA can be reduced by almost $70\%$.
As show in  Table~\ref{tab:sim_sbROCPCA}, BROC-PCA can provide not only  comparably accurate subspace recovery with the standard ROC-PCA, but  impressive gains in computational efficiency as especially when $p$ is large. For example, when $p=1000$, the computational cost of the standard ROC-PCA can be reduced by almost $70\%$.

\section{Real Data}
%\label{sec:simulation_rocpca}
\label{sec:realdata}

%\section{Numerical Experiments}
%%\label{sec:simulation_rocpca}
%\label{sec:realdata}
%
%%\input{simu}
%We performed extensive simulation studies to show the performance of ROC-PCA in the presence of r-Type/e-Type outliers.  Due to page limitations, the results are reported in the supplementary material.
%\subsection{Segmentation data}% from the UCI Machine Learning Repository  Among the $19$ features,

 We also applied ROC-PCA to analyze  a segmentation dataset collected by  Brodley~\citep{Bache+Lichman:2013} which contains features extracted from seven classes of hand-segmented images ({brickface, sky, foliage, cement, window, path} and {grass}). Each class has $330$ image regions, and for each image region $19$ features are provided (e.g., the contrast of vertically or horizontally adjacent pixels).  In this experiment, we randomly picked $90$ samples  from the {cement} class as normal observations and $10$ from the {foliage} class as outliers. %Because the third feature (number of pixels in each image region) is a constant, it is dropped in our experiment. In summary, we have $n=100$ with $1$-$90$ observations normal and $91$-$100$ abnormal.

In PCA applications, the adjusted variance \citep{shen2008sparse} is often used to assess the goodness of fit. We use a robust version, called {robust adjusted variance} (RAV), to take outliers into account. Let $\hat{\bsbV}_r$ be a robust  estimate  of the top $r$ loading vectors from data matrix $\bsbX$. The RAV explained by $\hat{\bsbV}_r$ is then defined as $\|\bsbX^0 \bsbP_{\hat{\bsbV}_r} \|_F^2/\|{\bsbX}^0\|_F^2$, where ${\bsbX}^0$ is a submatrix of $\bsbX$ containing  clean samples only. %, with the detected outliers removed.
% While the former aims at addressing the correlation and the lack of orthogonality among the estimated PC directions, the latter deals with the outlier effects and properly evaluates the variance explained by PCs in a robust manner.

\begin{figure}[!htb]
    \centering
    \includegraphics[height=0.32\textwidth,width=0.42\textwidth,,clip]{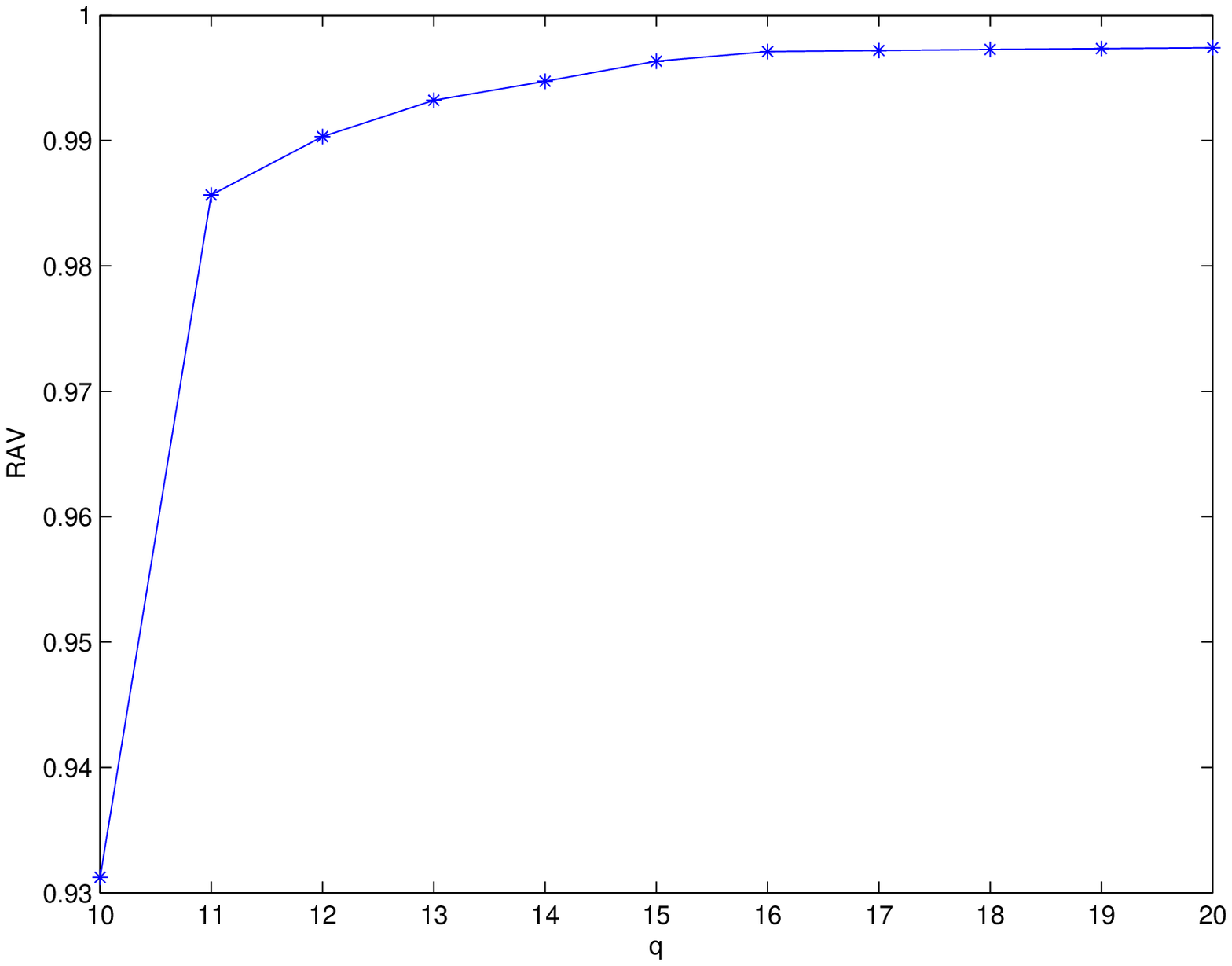}
    \includegraphics[height=0.32\textwidth,width=0.42\textwidth,,clip]{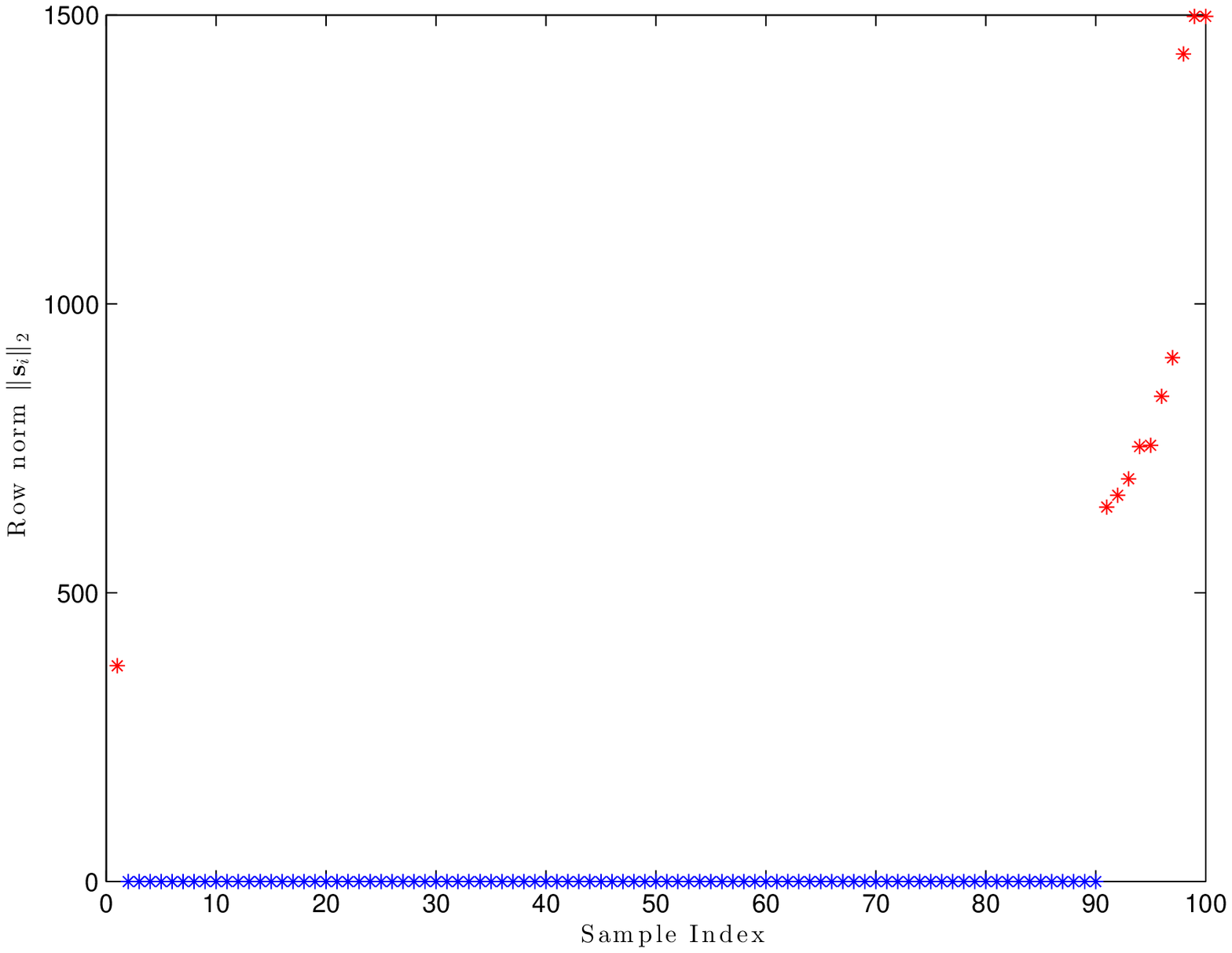}
    \caption{\small Left: RAV versus $q$ (with $r=3$) on the segmentation dataset. Right: Outlyingness plot:   row norms of $\hat{\bsbS}$. All nonzeros  are marked in red. The last 10 correspond to the foliage outliers, while the first one is from the cement class.}%the $\hat{\bsbS}$ estimate given by ROC-PCA with $q=11$ }
    % obtained from ROC-PCA with different values of $q$ on the segmentation data.}
    \label{fig:qtuningRealdata}
\end{figure}

We  ran the r-Type algorithm with $r=1,2,3$ and  $q$ decreasing from $20$ to $10$. Three principal components seem to be enough from the high RAV percentages in   Table \ref{tab:sim_realdata}. The left panel of Figure~\ref{fig:qtuningRealdata} shows a relatively big drop in  RAV  when  $q$ changes from $11$ to $10$.   ROC-PCA thus yielded 11 outliers rather than 10. To get some intuition,  we also plotted in the right panel %Figure~\ref{fig:imagesegmentOutlier11q}
the outlyingness for each observation, i.e.,  row norms of $\hat{\bsbS}$ (with $q=11$). The  last $10$ samples in the {foliage} class were successfully identified and most observations in the {cement} class have $\hat\bsbs_i$  exactly zero. Interestingly,   the first sample  pops up with a  large row norm of about ${400}$(!) in the outlyingness plot.  We examined  the  data  carefully and  verified this finding---for example, the 7th feature for the first observation takes value {375.1}, while the other $89$ samples in the same class show an average of only {1.8}.
%So it seems that Observation 1 does deviate from the remaining ones in the {cement} class.
{The pleasant finding  shows the power  of ROC-PCA in automatic outlier detection without much human intervention.}
% This demonstrates ROC-PCA's capability of automatic outlier identification.

%\begin{figure}[!htb]
%    \centering
%    \includegraphics[scale=0.4,clip]{./figures/imagesegmentOutlier11q.eps}
%    \caption{\small Outlyingness plot:   row norms of $\hat{\bsbS}$. All nonzeros  are marked in red. The last 10 correspond to the foliage outliers, while the first one is from the cement class.}%the $\hat{\bsbS}$ estimate given by ROC-PCA with $q=11$
%    \label{fig:imagesegmentOutlier11q}
%\end{figure}

We  also compared  ROC-PCA  with  PCA, PCP \citep{Candes2011},  S-PCA \citep{locantore1999robust}, RAPCA \citep{hubert2002fast} and ROBPCA \citep{hubert2005robpca} on the segmentation dataset.  Among  these methods only ROC-PCA and PCP  labeled outliers explicitly. The  outlying  entries detected by PCP scatter all  over the  matrix, and the zero/nonzero pattern provides little help in separating the foliage samples from the majority of  the cement data. Table~\ref{tab:sim_realdata} gives  the RAV rates explained by the top $1$-$3$ PCs. The  first six rows evaluated RAV on the  90  {cement}  samples. The true performance of ROC-PCA, as shown in the last row, was assessed  on  the 89 clean observations in consideration of its outlier detection. The improvement bought by   simultaneous   subspace recovery and outlier identification is significant.

\begin{table}[H]
\centering \caption{\small Explained variance in terms of RAV $\times 100$, by the top $1$-$3$ PCs. The RAV evaluation  of the first six rows is based on the cement class, and  ROC-PCA at the bottom is after removing all its   detected outliers.}
 \label{tab:sim_realdata} \footnotesize%\scriptsize
  \begin{tabular}{c | c  c  c  c }
  \multicolumn{5}{c}{} \\
          \hline

          \hline

\multicolumn{2}{c|}{}& 1 PC  & 2 PCs &  3 PCs  \\ \hline

\multicolumn{2}{l |}{PCA}
         &  5 &  20 & 66 \\
\multicolumn{2}{l |}{PCP}
         & 13 & 54 & 57 \\ %\hline
\multicolumn{2}{l |}{RAPCA}
         & 45 & 71 & 80  \\ %\hline
\multicolumn{2}{l |}{S-PCA}
         & 48 & 75 & 81  \\ %\hline
\multicolumn{2}{l |}{ROBPCA}
         & 48 & 75 & 82 \\ %\hline
\multicolumn{2}{l |}{$\text{ROC-PCA}^0$}
         & 48 & 76 & 82 \\ %\hline
         \hline
\multicolumn{2}{l |}{$\text{ROC-PCA}$}
         & \textbf{58} & \textbf{91} & \textbf{99} \\ \hline

          \hline
  \end{tabular}
\end{table}

\section{Conclusion}\label{sec:conclusion_rocpca}
%\section{Discussion}
We showed that PCA is sensitive to a type of  outliers which may not be easily revealed  in the original observation space; mathematically formulating these projected  outliers gave rise to the robust orthogonal complement PCA.  We showed that ROC-PCA  comes with a robust guarantee as generalized robust M-estimators, and provides ease in computation and regularization. Our theoretical analyses revealed the high breakdown point of ROC-PCA and established a non-asymptotic oracle inequality which can achieve the  minimax  error rate.

Some future research topics include   further studies of the  performance of the manifold optimization algorithm, as well as the development of faster algorithms in very high dimensions. Another interesting direction is to jointly investigate observation anomalies, which are of independent interest in many computer vision applications, and OC outliers (which can skew the PC subspace) to  provide a reinforced robustification of PCP.

%\section*{Supplementary Materials}
%The supplement materials provide a literature survey of robust PCA, an overall summary of the computational algorithm,   simulation studies and all technical details.
%%\pagebreak

\section{Proofs}
\label{sec:proofs}
\subsection{Proof of Theorem~\ref{mytheorem:Mestimate}}
\label{appendix:MThetaConnection}
Part (i): Let $(\hat{\bsbV}_{\perp}, \hat{\bsbmu}, \hat{\bsbS})$  be a  {coordinate-wise minimum} of the ROC-PCA problem~\eqref{eq:ewise} in the sense that fixing any two of $\hat{\bsbV}_{\perp}$, $\hat{\bsbmu}$, and  $\hat{\bsbS}$, the remaining one  is a (global) minimizer of \eqref{eq:ewise}.

The sub-problem for $\bsbVp$  is smooth on the Stiefel manifold.  Equipped with the canonical metric,
the Riemannian gradient, calculated in \eqref{eq:Rgradient},  vanishes at ${\hat \bsbV}_{\perp}$.  That is, $\bsbX^T(\bsbX\hat{\bsbV}_{\perp}-\bsb{1}\hat{\bsbmu}^T-\hat{\bsbS})= \hat{\bsbV}_{\perp}(\bsbX\hat{\bsbV}_{\perp}-\bsb{1}\hat{\bsbmu}^T-\hat{\bsbS})^T\bsbX\hat{\bsbV}_{\perp}$. The sub-problem for $\bsbS$ is not necessarily convex or smooth. However, based on Lemma 1 in~\cite{she2012iterative}, $\Theta(\bsbX\hat{\bsbV}_{\perp}-\frac{1}{n}\bsb{1}\bsb{1}^T(\bsbX\hat{\bsbV}_{\perp}-\hat{\bsbS});\lambda)$,  always gives a global minimizer for any $P$ coupled with $\Theta$ in the way of \eqref{eq:penaltyconstruction}; see Section \ref{sec:musopt_rocpca} in the text for details. Hence  $\hat{\bsbV}_{\perp}$, $\hat{\bsbmu}$ and $\hat{\bsbS}$ must satisfy the following equations
\begin{align}
%\begin{gather}
%\begin{eqnarray}
\bsbX^T(\bsbX\hat{\bsbV}_{\perp}-\bsb{1}\hat{\bsbmu}^T-\hat{\bsbS}) &= \hat{\bsbV}_{\perp}(\bsbX\hat{\bsbV}_{\perp}-\bsb{1}\hat{\bsbmu}^T-\hat{\bsbS})^T\bsbX\hat{\bsbV}_{\perp},\label{eq:coordminimum1}\\
\hat{\bsbS}&=\Theta(\bsbX\hat{\bsbV}_{\perp}-\frac{1}{n}\bsb{1}\bsb{1}^T(\bsbX\hat{\bsbV}_{\perp}-\hat{\bsbS});\lambda), \label{eq:coordminimum2}\\
\hat{\bsbmu}&=\frac{1}{n}(\bsbX\hat{\bsbV}_{\perp}-\hat{\bsb{S}})^T\bsb{1}. \label{eq:coordminimum3}
%\end{eqnarray}
%\end{gather}
\end{align}

Now, for $\psi(t) = t - \Theta(t;\lambda)$, we have
\begin{equation*}
\begin{array}{ll}
 & \psi(\bsbX\hat{\bsbV}_{\perp}-\bsb{1}\hat{\bsbmu}^T;\lambda) \\
=& \bsbX\hat{\bsbV}_{\perp}-\bsb{1}\hat{\bsbmu}^T-\Theta(\bsbX\hat{\bsbV}_{\perp}-\bsb{1}\hat{\bsbmu}^T;\lambda)\\
=& \bsbX\hat{\bsbV}_{\perp}-\bsb{1}\hat{\bsbmu}^T-\Theta(\bsbX\hat{\bsbV}_{\perp}-\frac{1}{n}\bsb{1}\bsb{1}^T(\bsbX\hat{\bsbV}_{\perp}-\hat{\bsbS});\lambda)\\
=& \bsbX\hat{\bsbV}_{\perp}-\bsb{1}\hat{\bsbmu}^T-\hat\bsbS,
\end{array}
\end{equation*}
where the last two equalities are by \eqref{eq:coordminimum3} and  \eqref{eq:coordminimum2}, respectively. It follows that

\begin{equation*}
\begin{array}{ll}
 & \bsbX^T\psi(\bsbX\hat{\bsbV}_{\perp}-\bsb{1}\bsbmu^T;\lambda)-\hat{\bsbV}_{\perp} (\psi(\bsbX\bsbVp-\bsb{1}\bsbmu^T;\lambda))^T\bsbX\hat{\bsbV}_{\perp}\\
=& \bsbX^T(\bsbX\hat{\bsbV}_{\perp}-\bsb{1}\hat{\bsbmu}^T-\hat{\bsbS}) - \hat{\bsbV}_{\perp}(\bsbX\hat{\bsbV}_{\perp}-\bsb{1}\hat{\bsbmu}^T-\hat{\bsbS})^T\bsbX\hat{\bsbV}_{\perp}\\
=& \bsb{0},
\end{array}
\end{equation*}
with the last equality due to \eqref{eq:coordminimum1}.
Moreover, we have \begin{equation*}
\begin{array}{ll}
 &\bsb{1}^T\psi(\bsbX\hat{\bsbV}_{\perp}-\bsb{1}\hat{\bsbmu}^T; \lambda)\\
=&\bsb{1}^T\psi(\bsbX\hat{\bsbV}_{\perp}-\frac{1}{n}\bsb{1}\bsb{1}^T(\bsbX\hat{\bsbV}_{\perp}-\hat{\bsbS}); \lambda) \ \ \mbox{(by \eqref{eq:coordminimum3})}\\
=&\bsb{1}^T(\bsbX\hat{\bsbV}_{\perp}-\frac{1}{n}\bsb{1}\bsb{1}^T(\bsbX\hat{\bsbV}_{\perp}-\hat{\bsbS})-\Theta(\bsbX\hat{\bsbV}_{\perp}-\frac{1}{n}\bsb{1}\bsb{1}^T(\bsbX\hat{\bsbV}_{\perp}-\hat{\bsbS});\lambda))\\
=&\bsb{1}^T(\bsbX\hat{\bsbV}_{\perp}-\frac{1}{n}\bsb{1}\bsb{1}^T(\bsbX\hat{\bsbV}_{\perp}-\hat{\bsbS})-\hat{\bsbS})\  \mbox{(by \eqref{eq:coordminimum2})}\\
=&\bsb{1}^T(\bsbI-\frac{1}{n}\bsb{1}\bsb{1}^T)\bsbX\hat{\bsbV}_{\perp}\\
=&\bsb{0}.
\end{array}
\end{equation*}
Therefore, $(\hat{\bsbV}_{\perp},\hat{\bsbmu})$ is also an $M$-estimate associated with $\psi$.

Part (ii): The proof   is straightforward and the details are omitted.

\subsection{Proof of Theorem~\ref{mytheorem:oracleerr}}
\label{appendix:oracleerr}

Throughout this proof, we use $C$, $c$, $L$ to denote universal constants. They are not necessarily the same at each occurrence.

Given any matrix $\bsbA$, we use $CS(\bsbA)$ and $RS(\bsbA)$ to denote its column space and row space, respectively, and   $\Proj_{\bsbA}$ to denote the orthogonal projection matrix onto   $CS(\bsbA)$, i.e., $\Proj_{\bsbA}=\bsbA(\bsbA^T\bsbA)^{+}\bsbA^T$, where $^+$ stands for  the Moore-Penrose pseudoinverse, while $\Proj_{\bsbA}^{\perp}$ denotes the projection onto the orthogonal complement of $CS(\bsbA)$. We also use such projection matrices to denote the associated subspaces by a bit abuse of notation. Given two matrices $\bsbA$ and $\bsbB$ of the same dimensions, their inner product is defined as $\langle \bsbA, \bsbB\rangle=tr(\bsbA^T \bsbB)$.

First, by the  definition of  ROC-PCA, we have
\begin{align}
\begin{split}
&\frac{np}{2} M(\hat \bsbA, \hat\bsbS, \hat\bsbV, \hat\bsbV_{\perp};\bsbA^*, \bsbS^*, \bsbV^*, \bsbV_{\perp}^*)    \\ \leq &\ \frac{np}{2} M( \bsbA, \bsbS, \bsbV, \bsbV_{\perp};\bsbA^*, \bsbS^*, \bsbV^*, \bsbV_{\perp}^*)  + \langle \bsbE, \hat\bsbA \hat\bsbV^T + \hat \bsbS \hat\bsbV_{\perp}^T - \bsbA\bsbV^T - \bsbS \bsbV_{\perp}^T \rangle.
\end{split}
\label{firstineq}
\end{align}
Define
$\mathcal J = \{i: \bsbS [i,:] \neq \bsb{0}\}$,
$\hat{\mathcal J} = \{i: \hat\bsbS [i,:] \neq \bsb{0}\}$, and  $\tilde {\mathcal  J} = \mathcal J \cup \hat{\mathcal J}$. For convenience, we denote  $\Proj_{\bsbI[:, \tilde {\mathcal J}]}$  by     $\Proj_{\tilde {\mathcal J}}$   and its  orthogonal complement by   $\Proj_{\tilde {\mathcal J}}^{\perp}$;    $\Proj_{\hat {\mathcal J}}$,  $\Proj_{ {\mathcal J}}$ and $\Proj_{ {\mathcal J}}^{\perp}$  are defined similarly.   Then
\begin{align*}
& \hat\bsbA \hat\bsbV^T + \hat \bsbS \hat\bsbV_{\perp}^T - \bsbA\bsbV^T - \bsbS \bsbV_{\perp}^T \\
= & (\hat\bsbA \hat\bsbV^T  +\hat \bsbS \hat\bsbV_{\perp}^T - \bsbA\bsbV^T ) \Proj_{\bsbV} + \Proj_{\tilde {\mathcal J}}^{\perp} \hat\bsbA \hat\bsbV^T     \Proj_{\bsbV_{\perp}} + \Proj_{ \tilde{\mathcal J}}( \hat\bsbA \hat\bsbV^T +\hat \bsbS \hat\bsbV_{\perp}^T    - \bsbS \bsbV_{\perp}^T) \Proj_{\bsbV_{\perp}} \\ =& (\hat\bsbA \hat\bsbV^T  +\hat \bsbS \hat\bsbV_{\perp}^T - \bsbA\bsbV^T ) \Proj_{\bsbV} + \Proj_{\tilde {\mathcal J}}^{\perp} \hat\bsbA \hat\bsbV^T \Proj_{\bsbV_{\perp}} + (\Proj_{ {\mathcal J}}\cap\Proj_{\tilde {\mathcal J}})( \hat\bsbA \hat\bsbV^T +\hat \bsbS \hat\bsbV_{\perp}^T    - \bsbS \bsbV_{\perp}^T) \Proj_{\bsbV_{\perp}} \\&+(\Proj_{\mathcal J}^{\perp}\cap\Proj_{\tilde {\mathcal J}}) ( \hat\bsbA \hat\bsbV^T +\hat \bsbS \hat\bsbV_{\perp}^T    - \bsbS \bsbV_{\perp}^T) \Proj_{\bsbV_{\perp}}\\
=& (\hat\bsbA \hat\bsbV^T  +\hat \bsbS \hat\bsbV_{\perp}^T - \bsbA\bsbV^T ) \Proj_{\bsbV} + \Proj_{\tilde {\mathcal J}}^{\perp} \hat\bsbA \hat\bsbV^T \Proj_{\bsbV_{\perp}} + \Proj_{ {\mathcal J}}( \hat\bsbA \hat\bsbV^T +\hat \bsbS \hat\bsbV_{\perp}^T    - \bsbS \bsbV_{\perp}^T) \Proj_{\bsbV_{\perp}} \\&+(\Proj_{\mathcal J}^{\perp}\cap\Proj_{\hat {\mathcal J}}) ( \hat\bsbA \hat\bsbV^T +\hat \bsbS \hat\bsbV_{\perp}^T    - \bsbS \bsbV_{\perp}^T) \Proj_{\bsbV_{\perp}}
\\\equiv & \bsbC_1 + \bsbC_2 + \bsbC_3 + \bsbC_4.
\end{align*}
Clearly,
$rank( \bsbC_1) \leq rank( \Proj_{\bsbV}) = r$,  $rank( \bsbC_2) \leq rank( \hat \bsbV) =r$, $rank( \bsbC_3) \leq rank( \Proj_{{\mathcal J}})  \leq  q$, $rank(\bsbC_4) \leq rank(\Proj_{\mathcal J}^{\perp}\cap\Proj_{\hat {\mathcal J}})\leq q$.  Due to the orthogonality between $\bsbC_1$, $\bsbC_2$,   $\bsbC_3$, and  $\bsbC_4$,  $\|\hat\bsbA \hat\bsbV^T + \hat \bsbS \hat\bsbV_{\perp}^T - \bsbA\bsbV^T - \bsbS \bsbV_{\perp}^T\|_F^2 =\sum_{l=1}^4 \|\bsbC_l\|_F^{2}$.
The last term of  \eqref{firstineq} now becomes
\begin{align}
\langle \bsbE, & \hat\bsbA \hat\bsbV^T + \hat \bsbS \hat\bsbV_{\perp}^T - \bsbA\bsbV^T - \bsbS \bsbV_{\perp}^T  \rangle = \sum_{l=1}^4 \langle \bsbE, \bsbC_l \rangle. \label{decomp4}
\end{align}
A lemma  will be introduced  to bound each  term on the right.
To make our conclusion  more general, we only require that $\bsbE$ is sub-Gaussian.
\begin{definition}\label{def:subgauss}
$\xi$ is called a sub-Gaussian random variable if there exist constants $C, c>0$ such that $\EP\{|\xi|\geq t\} \leq C e^{-c t^2}, \forall t>0$.  The  scale ($\psi_2$-norm) for  $\xi$ is defined as $\sigma( \xi) =
\inf \{\sigma>0: \EE\exp(\xi^2/\sigma^2) \leq 2\}$. $\bsbxi\in \mathbb R^p$ is called a sub-Gaussian random vector with scale  bounded by $\sigma$ if all one-dimensional marginals $\langle \bsbxi, \bsba \rangle$ are sub-Gaussian satisfying $\|\langle \bsbxi, \bsba \rangle\|_{\psi_2}\leq \sigma \|\bsba \|_2$, $\forall \bsba\in R^{p}$.   \ %\inf_{p\geq 1} p^{-1/2} (\EE|\bsbX|^p)^{1/p}$.
\end{definition}

Some examples include  Gaussian random variables and bounded random variables (such as Bernoulli).
The following lemma  assumes $\vect(\bsbE)$ is sub-Gaussian. % and is proved  in Appendix \ref{sec:proofoflemma1}.   %We write $xi\sim \mbox{sub-Gaussian}(\sigma^2)$ if $\xi$ is a  centered sub-Gaussian random variable with norm  bounded above by $\sigma$.

\begin{lemma}\label{emprocBnd}
Suppose $\bsbE \in \mathbb R^{n\times m}$ and $\vect(\bsbE)$ is  sub-Gaussian with mean zero and  $\psi_2$-norm bounded by $\sigma$.  %Suppose the rows of $\bsbE$ are independent    mean-centered sub-Gaussian random vectors with  $\psi_2$ norms bounded by $\sigma$.
(i) Given $\bsbX\in \mathbb R^{n\times p}$, $1\leq J\leq p$, $1\leq r \leq J\wedge m$, define $\Gamma_{J, r}^{\bsbX} = \{\bsbA\in \mathbb R^{n\times m}: \|\bsbA\|_F\leq 1, \mbox{rank}(\bsbA)\le r, CS(\bsbA) \subset \Proj_{\bsbX_{\mathcal J}}) \mbox{ for some } \mathcal J: | \mathcal J|=J\}$ with $\bsbX_{\mathcal J}$ denoting the submatrix consisting of the columns of $\bsbX$ indexed by $\mathcal J$. Let
  $P_o^{\prime}(J, r) = \sigma^2 \{ [\mbox{rank}(\bsbX)\wedge  J +m - r] r + \log {p \choose J}\}$. Then for any $t\geq 0$,
\begin{align}
\EP \left(\sup_{\bsbA \in \Gamma_{J,r}^{\bsbX}} \langle \bsbE, \bsbA \rangle \geq t \sigma +   \sqrt{L \cdot P_o^{\prime}(J,r)}\right) \leq C\exp(- ct^2),
\end{align}
where $L, C, c>0$ are universal constants.
(ii) Given $\bsbX\in \mathbb R^{n\times p}$, $1\leq J, J'\leq p$, $1\leq r \leq J\wedge m$, define $\Gamma_{J,J', r}^{\bsbX} = \{\bsbA\in \mathbb R^{n\times m}: \|\bsbA\|_F\leq 1, rank(\bsbA) \leq r,  CS(\bsbA) \subset \Proj_{\bsbX_{\mathcal J}}^{\perp} \cap \Proj_{\bsbX_{\mathcal J'}}   \mbox{ for some }  {\mathcal J}, {\mathcal J}' \subset [p] \mbox{ satisfying }  | {\mathcal J}'|=J',  | {\mathcal J}|=J \}$. Let $P_o''(J,J', r) = \sigma^2 \{(rank(\bsbX)\wedge  J' \wedge (p- J)) r  +(m - r) r + \log {p\choose J} + \log {p\choose J'}\}$.
Then for any $t\geq 0$,
\begin{align}
\EP \left(\sup_{\bsbA \in \Gamma_{J, J',r}^{\bsbX}} \langle \bsbE, \bsbA \rangle \geq t \sigma +   \sqrt{L \cdot P_o''(J,J',r)}\right) \leq C\exp(- ct^2),
\end{align}
where $L, C, c>0$ are universal constants.
\end{lemma}

The proof of the lemma follows similar  lines of the proof of Lemma 4 in \cite{SheSelrrr} and is omitted.

 Noticing that $\bsbC_1, \bsbC_2 \in  \Gamma_{n,r}^{\bsbI}$, $\bsbC_3 \in  \Gamma_{q, d }^{\bsbI}$, $\bsbC_4\in \Gamma_{q,q, d}^{\bsbI}$, we can apply the lemma to bound \eqref{decomp4}.  Take   $\bsbC_3$ as an example. Letting $P_o^{(3)} (q, d) = \sigma^2 \{q d + (p-d)d +  \log {n \choose q}\}$,  we have
$
  \langle \bsbE, \bsbC_3 \rangle - \frac{1}{a} \|\bsbC_3 \|_F^2 -  b L P_o^{(3)}(q,  d)
\leq
\|\bsbC_3\|_F  \langle \bsbE, \bsbC_3 / \|\bsbC_3\|_F\rangle - 2\sqrt{\frac{  b}{a}} \|\bsbC_3\|_F \sqrt{L P_o^{(3)}( q,  d)}
%\leq
$,
which is further bounded by
$\frac{1}{a^{\prime}} \|\bsbC_3\|_F^2 +  \frac{a^{\prime}}{4}
\sup_{1\leq  q \leq n, 1\leq d \leq p} \left( \sup_{\bsbA\in \Gamma_{q, d}^{\bsbI}} \langle \bsbE, \bsbA \rangle - 2\sqrt{\frac{  b}{a}} \sqrt{L P_o^{(3)}(q, d)} \right)_+^2$
for any $a, b, a^{\prime} >0$.  Denote the last term by $\frac{a^{\prime}}{4} R^2$ with $R^2=\sup_{1\leq  q \leq n, 1\leq d \leq p} R_{ q, d}$. From Part (i) of Lemma \ref{emprocBnd},  $\EE R^2 \leq C \sigma^2$ if we choose  $b>4 a$. Indeed,
\begin{align*}
&\quad\EP(R\geq t \sigma )\\
 &\leq  \sum_{ q=1}^n \sum_{d=1}^{p} \EP(R_{ q, d}\geq t\sigma )\\
 &\leq  \sum_{q=1}^n \sum_{d=1}^{p}  \EP (\sup_{\bsbA \in \Gamma_{ q, d}^{\bsbI}}  \langle \bsbE, \bsbA \rangle -\sqrt{L  P_o^{(3)}( q, d)} \geq t \sigma +   (2\sqrt{{b}/{a}}-1)\sqrt{L  P_o^{(3)}( q, d)})
\\
& \leq  \sum_{ q=1}^n \sum_{d=1}^{p} C \exp(-c t^2) \exp\left\{- c \left((2\sqrt{{b}/{a}}-1)^2 L \cdot P_o^{(3)}( q, d)/\sigma^2\right)\right\}
 \leq  C^{\prime} \exp(-c t^2).
\end{align*}
The other terms $\langle \bsbE, \bsbC_1 \rangle, \langle \bsbE, \bsbC_2 \rangle$, and  $\langle \bsbE, \bsbC_4 \rangle$  can be similarly handled by Lemma \ref{emprocBnd}.

In summary, letting  we obtain a bound for  \eqref{decomp4}  as follows\begin{align*}
&\EE \langle \bsbE, \hat\bsbA \hat\bsbV^T + \hat \bsbS \hat\bsbV_{\perp}^T - \bsbA\bsbV^T - \bsbS \bsbV_{\perp}^T \rangle \\
 \leq &  \EE \{(\frac{1}{a}+\frac{1}{a^{\prime}}) \|\hat\bsbA \hat\bsbV^T + \hat \bsbS \hat\bsbV_{\perp}^T - \bsbA\bsbV^T - \bsbS \bsbV_{\perp}^T\|_F^2\} +C \sigma^2+   \\
& +  b L \sigma^2\{2n r  + 2(p-r)r +qd+(p-d)d +qd + (p-d)d +  3  \log {n \choose J}\} \\
\leq & \EE\{(\frac{1}{a}+\frac{1}{a^{\prime}})(1+b^{\prime}) \|\bsbA \bsbV^T +  \bsbS  \bsbV_{\perp}^{ T} - \bsbA^*\bsbV^{* T} - \bsbS^* \bsbV_{\perp}^{* T} \|_F^2 \\
 & +  (\frac{1}{a}+\frac{1}{a^{\prime}})(1+\frac{1}{b^{\prime}}) \|\hat\bsbA \hat\bsbV^T + \hat \bsbS \hat \bsbV_{\perp}^{ T} - \bsbA^*\bsbV^{* T} - \bsbS^* \bsbV_{\perp}^{* T} \|_F^2\} \\
&  +  b C L \sigma^2\{qp + p r + nr + q\log (e n /q)\},
\end{align*}
for any $a, b, a^{\prime}, b^{\prime}$ that are positive and  satisfy $b> 4a$.
Choosing   $(\frac{1}{a}+\frac{1}{a^{\prime}})(1+\frac{1}{b^{\prime}})<\frac{1}{2}$ and  $b>4a$ (e.g., $a=a'=8$, $b'=2$, $b=33$),   we  get
$
n p \EE M(\hat \bsbA, \hat\bsbS, \hat\bsbV, \hat\bsbV_{\perp};\bsbA^*, \bsbS^*, \bsbV^*, \bsbV_{\perp}^*)  \leq  C n p M( \bsbA, \bsbS, \bsbV, \bsbV_{\perp};\bsbA^*, \bsbS^*, \bsbV^*, \bsbV_{\perp}^*)  + C\sigma^2 + np P_o(q, r;  n, p, \sigma^2).
$

\subsection{Proof of Theorem \ref{th_minimax}}
Let $\bsbB(\bsbA, \bsbS, \bsbV, \bsbV_{\perp}) = \bsbA \bsbV^T + \bsbS \bsbV_{\perp}^T$. With a bit abuse of notation, the set of $\bsbB(\bsbA, \bsbS, \bsbV, \bsbV_{\perp})$ for all $ (\bsbA, \bsbS, \bsbV, \bsbV_{\perp}) \in \mathcal S(r,q)$ is still denoted by $\mathcal S(r,q)$.
%The proof is based on the general reduction scheme  in Chapter 2 of \cite{tsybakov2009introduction}. The key is to design proper {least favorable} signals  in different situations.

\textit{Case (i)} $(n+p)r < q d+q \log (e n /q) $.
Define a signal subclass ${\mathcal B}^1(r, q)=\{\bsbB(\bsbA, \bsbS, \bsbV, \bsbV_{\perp}): \bsbA= \bsb{0},  \bsbS\in {\mathcal B}_S^1(r, q), \bsbV = \bsbI[:, 1:r], \bsbV_{\perp}= \bsbI[:, (r+1):p] \}$, where
\begin{align*}
{\mathcal B}_S^1(r, q)&=\{ \bsbS \in \mathbb R^{n\times d}:  \bsbS =[\bsbS^l, \bsbS^r], \bsbS^l= [\bsbs_{1}^l, \ldots, \bsbs_n^l]^T\in \mathbb R^{n\times [d/2]} \mbox{ with } \bsbs_i^l \in\{ \bsb{0}, \gamma R \cdot\bsb{1}\}, \\
&\bsbS^r = [s_{i,j}^r] \in \mathbb R^{n\times (d-[d/2])}, s_{i,j}^r=0 \mbox{ if }  \bsbs_i^l=\bsb{0}, \mbox{ and } \in \{0, \gamma R \}\mbox{ otherwise}, \mbox{ and } \|\bsbS\|_{2,0}\leq q \}.
\end{align*}
Here  $R= {\sigma}  (1+ \frac{\log (e n /q)}{d})^{1/2} $ and $\gamma>0$ is a small constant to be chosen later.
% (based on Theorem 2.5 of \cite{tsybakov2009introduction}).
 Clearly,  ${\mathcal B}^{1}(r, q) \subset\mathcal{S}(r,q) $.
By Stirling's approximation, $\log |\mathcal B^{1}(q)|\ge \log { n \choose q} + \log 2 ^{qd/2}\geq  q\log (n/q)+(qd/2) \log 2\geq c \{q \log ( en/q) + qd\}$ for some universal constant $c$.
Let $\rho(\bsbS_1, \bsbS_2)=\|\vect(\bsbS_1) - \vect(\bsbS_2)\|_0$ be the Hamming distance. Applying  Lemma 8.3 of   \cite{Rigollet11} row-wise, followed by the  Varshamov-Gilbert bound (cf. Lemma 2.9 in \cite{tsybakov2009introduction}),  there exists  a subset ${\mathcal B}_S^{10}(r, q)\subset {\mathcal B}_S^{1}(r, q)$ such that
\begin{eqnarray}
\log | {\mathcal B}_S^{10}(r, q)| \geq c_1 (q\log ( e n/q)+ qd )
\mbox{ and }
\rho(\bsbS_1, \bsbS_2) \geq c_2 q d , \forall \bsbS_1, \bsbS_{2} \in {\mathcal B}_S^{10},  \bsbS_1\neq \bsbS_2
\end{eqnarray}
for some universal constants $c_1, c_2>0$. Let ${\mathcal B}^{10}(r, q)=\{\bsbB(\bsbA, \bsbS, \bsbV, \bsbV_{\perp}): \bsbA= \bsb{0},  \bsbS\in {\mathcal B}_S^{10}(r, q), \bsbV = \bsbI[:, 1:r], \bsbV_{\perp}= \bsbI[:, (r+1):p] \}$. Then  for any $\bsbB_1, \bsbB_{2} \in \mathcal B^{10}(r, q)$,  $\bsbB_1\neq \bsbB_2$,
\begin{align}\| \bsbB_1 - \bsbB_2\|_F^2 = \gamma^2 R^2 \rho(\bsbB_1, \bsbB_2) \geq c_2 \gamma^2 R^2 q d. \label{separationLBound}
\end{align}

For Gaussian models, the  Kullback-Leibler divergence of $\mathcal {M N}(  \bsbB_2,\sigma^2 \bsb{I}\otimes\bsbI)$ (denoted by $P_{\bsbB_2}$) from $\mathcal {MN}(  \bsbB_1,\sigma^2 \bsb{I}\otimes \bsbI)$ (denoted by $P_{\bsbB_1}$) is $\mathcal K(\mathcal P_{\bsbB_1}, \mathcal P_{\bsbB_2}) = \frac{1}{2\sigma^2} \| \bsbB_1 -   \bsbB_2 \|_F^2$. Let $P_{\bsb{0}}$ be $\mathcal {M N}(\bsb{0}, \sigma^2 \bsb{I}\otimes \bsbI)$. Then,
for any $\bsbB \in {\mathcal B}^{1}(r, q) $, we have
\begin{align*}
\mathcal K(P_{\bsb{0}}, P_{\bsbB}) =\frac{1}{2\sigma^2} \| \bsb{0} -   \bsbS \|_F^2 \leq \frac{\gamma^2}{\sigma^2} R^2 qd
\end{align*}
and so
\begin{align}
\frac{1}{|\mathcal B^{10}|}\sum_{\bsbB\in \mathcal B^{10}(r,q)} \mathcal K(P_{\bsb{0}}, P_{\bsbB})\leq \gamma^2 (qd+q
\log(en/q)). \label{KLUBound}
\end{align}

Combining  \eqref{separationLBound} and \eqref{KLUBound} and choosing a sufficiently small value for  $\gamma$, we can apply Theorem 2.7 of \cite{tsybakov2009introduction} to get the desired  lower bound.

\textit{Case (ii)} $(n +p ) r \geq q d+  q\log (e n /q)$.
Consider a signal  subclass \begin{align*}
{\mathcal B}^2 (r)=\{\bsbB = [b_{i,j}]\in \mathbb R^{n\times p}:   & \ b_{i,j}=0 \mbox{ or } \gamma \sigma \mbox{ if }  (i,j)\in[n]\times[ r/2]\cup [r/2]\times[ p] \\ & \mbox{ and }   b_{i,j}=0 \mbox{ otherwise} \}.
\end{align*}
where  $\gamma>0$ is a small constant to be chosen later.
Then it is not difficult to see that   $|{\mathcal B}^2 (r)|= 2^{(n+p-r/2)r}$, $\mathcal B^2 (r)\subset \mathcal B(r, q)$, and $r(\bsbB_1 - \bsbB_2)\leq r$, $\forall \bsbB_1, \bsbB_2 \in \mathcal B^2(r)$. Also, since $r\le n\wedge p$, $(n+p-r/2)r\geq c(n+p)r$ for some universal constant $c$.

Let $\rho(\bsbB_1, \bsbB_2)=\|\vect(\bsbB_1) - \vect(\bsbB_2)\|_0$ be the Hamming distance. By the  Varshamov-Gilbert bound,
there exists  a subset ${\mathcal B}^{20}(r)\subset {\mathcal B}^{2}(r)$ such that
\begin{eqnarray*}
\log | {\mathcal B}^{20}(r)| \geq c_1 r(n+p), \mbox{ and }
\rho(\bsbB_1, \bsbB_2) \geq c_2 r(n+p), \forall \bsbB_1, \bsbB_{2} \in \mathcal B^{20},  \bsbB_1\neq \bsbB_2
\end{eqnarray*}
for some universal constants $c_1, c_2>0$.
Therefore,  $\| \bsbB_1 - \bsbB_2\|_F^2 = \gamma^2 \sigma^2 \rho(\bsbB_1, \bsbB_2) \geq c_2 \gamma^2 \sigma^2 (n+p) r$,
 for any $\bsbB_1, \bsbB_{2} \in \mathcal B^{20}$,  $\bsbB_1\neq \bsbB_2$. The afterward treatment follows the same lines as in (i) and the details are omitted.

\bibliographystyle{apalike}
\bibliography{ROCPCA}
%}

%\newpage
%\appendix
%\appendixpage
%%%\addappheadtotoc

\end{document}